\documentclass[nojss]{jss}
\usepackage{verbatim}
\usepackage{amsfonts}
\usepackage{amsmath}
\usepackage{natbib}
\usepackage{color}
\usepackage{graphicx}

\definecolor{red}   {RGB}{180,0,0}
\definecolor{gray10}{rgb}{0.1,0.1,0.1}
\definecolor{gray20}{rgb}{0.2,0.2,0.2}
\definecolor{gray30}{rgb}{0.3,0.3,0.3}
\definecolor{gray40}{rgb}{0.4,0.4,0.4}
\definecolor{gray50}{rgb}{0.5,0.5,0.5}
\definecolor{gray60}{rgb}{0.6,0.6,0.6}
\definecolor{gray80}{rgb}{0.8,0.8,0.8}
\definecolor{gray90}{rgb}{0.9,0.9,.9}
\definecolor{gray95}{rgb}{0.95,0.95,.95}
\definecolor{gray96}{rgb}{0.96,0.96,.96}
\definecolor{sgGreen} {RGB}{20, 180, 50}

\newcommand{\overbar}[1]{\mkern 1.5mu\overline{\mkern-1.5mu#1\mkern-1.5mu}\mkern 1.5mu}

\author{Victoria Lin\thanks{V. Lin and S. McGrath made equal contributions.} \\ School of Computer \\ Science \\ Carnegie Mellon University
    \And Sean McGrath\footnotemark[1] \\ Harvard T. H. Chan \\ School of \\ Public Health
	\And Zilu Zhang \\ Harvard Medical School \\ \ Dana Farber \\  Cancer Institute
	\AND Lucia C. Petito \\ Feinberg School \\of Medicine \\ Northwestern University
	\And Roger W. Logan \\ Harvard T. H. Chan \\ School of \\ Public Health
	\And Miguel A. Hern\'an\thanks{M.A. Hern\'an and J.G. Young made equal contributions. } \\ Harvard T. H. Chan \\ School of \\ Public Health
	\And Jessica G. Young\footnotemark[2] \\ Harvard Medical School\\  \ Harvard Pilgrim \\  Health Care Institute}
\title{\pkg{gfoRmula}: An \proglang{R} package for estimating effects of general time-varying treatment interventions via the parametric g-formula}

\Plainauthor{Victoria Lin and Sean McGrath, co-first authors} 
\Plaintitle{gfoRmula: An R package for estimating effects of general time-varying treatment interventions via the parametric g-formula} 
\Shorttitle{\pkg{gfoRmula}: The parametric g-formula in \proglang{R}} 

\Abstract{
  Researchers are often interested in using longitudinal data to estimate the causal effects of hypothetical time-varying treatment interventions on the mean or risk of a future outcome. Standard regression/conditioning methods for confounding control generally fail to recover causal effects when time-varying confounders are themselves affected by past treatment. In such settings, estimators derived from Robins's g-formula may recover time-varying treatment effects provided sufficient covariates are measured to control confounding by unmeasured risk factors. The package \pkg{gfoRmula} implements in \proglang{R} one such estimator: the parametric g-formula. This estimator easily adapts to binary or continuous time-varying treatments as well as contrasts defined by static or dynamic, deterministic or random treatment interventions, as well as interventions that depend on the natural value of treatment. The package accommodates survival outcomes as well as binary or continuous end of follow-up outcomes. For survival outcomes, the package has different options for handling competing events. This paper describes the \pkg{gfoRmula} package, along with motivating background, features, and examples.
}
\Plainkeywords{g-formula, longitudinal data, causal inference, R} 

\Address{
  Miguel Hern\'an \\
  Program on Causal Inference \\
  Harvard T. H. Chan School of Public Health \\
  677 Huntington Ave, Boston, MA 02115 \\
  E-mail: \email{mhernan@hsph.harvard.edu} \\
}

\begin{document}


\section{Introduction}
Researchers are often interested in using longitudinal data to estimate the causal effects of hypothetical time-varying treatment interventions (equivalently, \textsl{strategies} or \textsl{rules}) on the mean or risk of a future outcome in a study population. Standard regression or conditioning methods for confounding control generally fail to recover such causal effects when time-varying confounders are themselves affected by past treatment \citep{chap207}.  For example, in studies of the effect of different time-varying antiretroviral treatment strategies on long-term mortality risk in HIV-infected patients, CD4 cell count is a time-varying confounder; i.e., CD4 cell count at follow-up time $k$ is a risk factor for both future treatment initiation and future mortality.  In addition, CD4 cell count at follow-up time $k$ is, itself, affected by whether treatment has been previously initiated.  As another example, in studies of the effect of different time-varying interventions on daily minutes of physical activity on long-term coronary heart disease (CHD) risk in a healthy population, body mass index (BMI) is a time-varying confounder; i.e., BMI at follow-up time $k$ is a risk factor for future inactivity and future CHD. Also, BMI at time $k$ is, itself, affected by an individual's past physical activity. 

In such settings, alternative estimators derived from Robins's \textsl{g-formula} may recover effects of time-varying treatment interventions under untestable assumptions, including that sufficient covariates are measured to control confounding by unmeasured risk factors \citep{robinsfail}. The package \pkg{gfoRmula} implements in \proglang{R} one such estimator: \textsl{the parametric g-formula}, also known as \textsl{parametric g-computation} or \textsl{the plug-in g-formula} \citep{gformula_Rpackage, causalbook}.  In general, the g-formula characterized by a user-specified treatment intervention is a high-dimensional sum or integral over all observed treatment and confounder histories.  This sum is over a function of (i) the observed outcome mean (or, in survival settings, time-varying hazards) conditional on each observed treatment and confounder history, (ii) the observed joint distribution of the confounders at each time $k$ conditional on each observed treatment and confounder history, and (iii) an intervention density defining the time-varying user-specified treatment rule.  The parametric g-formula estimates this function in realistic high-dimensional settings via a Monte Carlo simulation that relies on consistent estimates of the quantities (i) and (ii), as well as the user-specified intervention density.

The \pkg{gfoRmula} package accommodates general user-specified time-varying treatment interventions which may be \textsl{static} or \textsl{dynamic} \citep{dynamicmsm1,dynamicmsm2,dynamicmsm3,historyadjmsm,murphy01,cainijb,pgformyoung} and, further, \textsl{deterministic} or \textsl{random} \citep{shiftint,haneuse,youngthreshold,repintyoung,kennedy}.  The algorithm implemented by the package can additionally estimate the extended g-formula of \cite{WHOchap} which, under a stronger no unmeasured confounding assumption, may identify effects of treatment assignment rules that depend on the \textsl{natural value of treatment} at $k$; the value of treatment that would have been observed at $k$ were the intervention discontinued right before $k$ \citep{swigs}.  In addition to the quantities (i) and (ii), the parametric extended g-formula relies on an estimate of the observed distribution of treatment at each time $k$ conditional on past treatment and confounder history. 

The package also allows: 1) binary (e.g. treat versus do not treat) or continuous/multi-level (e.g. dose, daily minutes of physical activity) time-varying treatments; 2) different types of outcomes (survival or continuous/binary end of follow-up); 3) data with competing events (survival outcomes) or truncation by death (end of follow-up outcomes) and loss to follow-up and other types of censoring events; 4) different options for handling competing events in the case of survival outcomes; 5) a random measurement/visit process; 6) joint interventions on multiple treatments; and 7) general incorporation of \textsl{a priori} knowledge of the data structure. The \pkg{gfoRmula} \proglang{R} package adapts many of the capabilities of the \pkg{GFORMULA} \proglang{SAS} macro to implement the parametric g-formula \citep{documentation}. However, unlike the \proglang{SAS} macro, the \pkg{gfoRmula} \proglang{R} package more easily allows users to incorporate their own helper functions for estimating (i) and (ii) beyond automated options included within the package.

The structure of the article is as follows: In Section \ref{background}, we review the longitudinal data structure of interest, the g-formula and identifying assumptions that give equivalence between this function and a counterfactual mean or risk under a user-specified time-varying treatment strategy.  In Section \ref{algorithm}, we review the parametric g-formula estimation algorithm.  In Section \ref{package}, we describe input data set requirements and core features of the package.  In Section \ref{examples}, we give various code examples.  In Section \ref{additional}, we describe additional, more advanced package features.  Finally, in Section \ref{discussion}, we provide a discussion.

\section{Background}\label{background}

\subsection{Observed data structure}
We consider a longitudinal study where measurements of treatment(s), and covariates are regularly updated on each of $i=1,\ldots ,n$ subjects over a specified follow-up period. Let $k=0,\ldots ,K+1$ denote fixed follow-up intervals (e.g. months) with baseline measurements taken in interval $k=0$ and interval $K+1$ corresponding to the specified end of follow-up (e.g. 60 months). 

In each interval $k,$ assume the following are measured: Let $A_{k}$ be a treatment variable (or vector of treatment variables) measured in interval $k$ (e.g. an indicator of antiretroviral treatment initiation by interval $k$; minutes of physical activity in interval $k$) and $%
L_{k}$ a vector of time-varying covariates  (e.g., CD4 cell count, BMI) assumed to precede $A_k$ with $L_{0}$ possibly 
including time-fixed baseline covariates (e.g., race, baseline age).  In clinical cohorts, including studies based on electronic medical records, $L_k$ may be defined in terms of last measured values of a covariate relative to interval $k$.  For example, in a clinical cohort of HIV-infected patients, CD4 cell count is not measured every month.  It is only measured in a month when the patient comes for a visit.  In this case $L_k$ may be defined in terms of the last measured value of CD4 cell count relative to month $k$.  Further, $L_k$ may contain an indicator of whether that last measured value is current or not (i.e. if a visit at $k$ has occurred) \citep{obsplans}.  

Let $Y\equiv Y_{K+1}$ denote the outcome of interest. $Y$ may correspond to either a fixed (continuous or binary) end of follow-up outcome \textsl{in} interval $K+1$ (e.g. blood pressure \textsl{in} interval $K+1$; an indicator of obesity \textsl{in} interval $K+1$) or a survival outcome; i.e., an indicator of failure from an event of interest \textsl{by} $K+1$ (e.g. an indicator of stroke \textsl{by} $K+1$).  For survival outcomes, we implicitly include in $L_k$  an indicator of failure from the event of interest by earlier $k<K+1$ ($Y_k$).  Past values of outcomes at the end of follow-up may also be components of $L_k$ (e.g. when the outcome is obesity status at $K+1$, obesity at any $k<K+1$ may be a component of $L_k$).  Note that ``end of follow-up'' may correspond to any user-specified follow-up of interest for which data is available and need not correspond to the administrative end of the study (e.g. the user may specify $K+1$ as 10 month follow-up even if the administrative end of the study was 60 months).  

Let $C_k$ denote an indicator of censoring and $D_k$ an indicator of a competing event (for a survival outcome) or truncation by death (for an end of follow-up outcome) by interval $k$, respectively. When a survival outcome is not subject to competing events (e.g. all-cause mortality), we define $D_k\equiv 0$ for all $k$. For end of follow-up outcomes, the user must define $D_k$ as an implicit component of $C_k$ (i.e. as a censoring event).  For survival outcomes, the user may choose to define $D_k$ as an implicit component of $C_k$ or $L_k$.  Under the latter choice, competing events are not treated as censoring events.  See \cite{crpaper} as well as below regarding how this choice impacts the interpretation of the analysis and estimation algorithms. 

We denote
the history of a random variable using overbars; for example, $\overline{A}%
_{k}=(A_{0},\ldots ,A_{k})$ is the observed treatment history through
interval $k$. By notational convention, we set $\overline{L}_{-1}$ and $%
\overline{A}_{-1}$ to be identically $0$. 

\subsection{Causal effect definitions}\label{estimands}
In each interval $k$, we generally define an intervention on $A_k$ that may, at most, depend on past measured variables as a random draw
from an intervention density $f^{int}(a_{k}|\overline{l}_{k},\overline{a}%
_{k-1},\overline{C}_k=0)$ with $(\overline{a}_k,\overline{l}_k)$ a possible realization of $(\overline{A}_{k},\overline{L}_{k})$.  

We classify an intervention $f^{int}(a_{k}|\overline{l}_{k},\overline{a}%
_{k-1},\overline{C}_k=0)$ on $A_k$ as \emph{deterministic} if $%
f^{int}(a_{k}|\overline{l}_{k},\overline{a}_{k-1},\overline{Y}_{k}=0)$ may
only equal zero or one for all $k $ and $(\overline{a}_{k},\overline{l}_{k})$%
; otherwise we classify it as \emph{random}. We may further classify the
intervention as \emph{static} if $f^{int}(a_{k}|\overline{l}_{k},\overline{a}%
_{k-1},\overline{C}_{k}=0)$ does not depend on $\overline{l}_{k}$ for any $k$; otherwise we classify it as \emph{dynamic}.  See \cite{youngthreshold} and \cite{repintyoung}, as well as Sections \ref{strategies} and \ref{customint}, for examples. 

Beyond interventions of the form $f^{int}(a_{k}|\overline{l}_{k},\overline{a}%
_{k-1},\overline{C}_k=0)$, we will also define an intervention on $A_k$ that, in addition to past measured variables, depends on the natural value of treatment at $k$.  We define such an intervention as a random draw from an alternative intervention density $f^{d}(a_{k}|a^{*}_k,\overline{l}_{k},\overline{a}%
_{k-1},\overline{C}_k=0)$ where $a^{*}_k$ is any value in the support of $A_k$ which, in the observational study, coincides with the natural value of treatment at $k$ \citep{swigs,youngthreshold}.  

Our goal is to estimate the causal effect on the mean of $Y$ (which corresponds to the risk of the event of interest by $K+1$ in the case of survival outcomes) had, contrary to fact we implemented two different hypothetical interventions on $A_k$ at all $k=0,\ldots,K$ in the study population where these interventions may be any user-specified choices of either  $f^{int}(a_{k}|\overline{l}_{k},\overline{a}%
_{k-1},\overline{C}_k=0)$ or $f^{d}(a_{k}|a^{*}_k,\overline{l}_{k},\overline{a}%
_{k-1},\overline{C}_k=0)$.  Implicit in the definition of all interventions is ``abolish censoring throughout follow-up'' or ``Set $C_k=0$ at all $k$''.  By this, how the user chooses to define $C_k$ will impact the interpretation of the estimand \citep{causalbook}.  For survival outcomes, when $D_k$ is defined as an implicit component of $C_k$, this effect is a special case of a direct effect that does not capture any treatment effect on the competing event; otherwise it is a special case of a total effect that may capture these effects \citep{crpaper}.   

\subsection{Identifying assumptions and the g-formula}\label{assumptions}
Let $g=(g_{0},...g_{K})$ be a \emph{deterministic} regime, strategy, or intervention (static or dynamic) on 
$A_k$ that depends, at most, on the measured covariate and treatment history, characterized by the intervention density $f^{int}(a_{k}|\overline{l}%
_{k},\overline{a}_{k-1}^{g},\overline{C}_{k}=0)=I(a_k=a_{k}^{g})$ that also eliminates censoring, where $a_{s}^{g}=g_{s}\left( \overline{\ell }_{s},\overline{a}%
_{s-1}^{g}\right) $ is any component of $\overline{a}_{k}^{g}=(a_{0}^{g},%
\ldots ,a_{k}^{g})$ and $\overline{a}_{s}^{g}$ is recursively defined by the
function $g_{s}$ of $(\overline{l}_{s},\overline{a}_{s-1}^{g})$, $s=0,\ldots
,k$.   

Define $Y^{g}$ and $\overline{L}%
_{K}^{g}$ as the outcome and covariate
histories, respectively, for an individual in the study population had, possibly contrary to fact, his/her treatment been assigned according to a deterministic regime $g$. Further, let $\mathcal{G}$ be the \emph{set of all deterministic interventions $g$}
on $A_k$ (both static and dynamic).  Following \cite{robinsfail}, the mean of $Y$ had all subjects been
assigned treatment according to $f^{int}(a_{k}|\overline{l}_{k},\overline{a}%
_{k-1},\overline{C}_{k}=0)$ and had censoring been eliminated is equivalent to $\sum_{g\in\mathcal{G}} w(g)\mbox{E}[%
Y^{g}]$ where $w(g)$ is a $g-$specific weight defined in terms of the choice of $f^{int}(a_{k}|\overline{l}_{k},\overline{a%
}_{k-1},\overline{Y}_{k}=0)$.  See also the appendix of \cite{youngthreshold}. 

We now define
three $g$-specific identifying conditions:

\begin{enumerate}

\item Exchangeability: For all $k=0,\ldots,K$
\begin{equation}
Y^{g} \coprod (A_{k},C_{k+1})|\overline{L}_{k}=%
\overline{l}_{k},\overline{A}_{k-1}=\overline{a}_{k-1}^{g},C_{k}=0
\label{exchstrong}
\end{equation}
The exchangeability condition (\ref{exchstrong}) is expected to hold in an
experiment where the treatment $A_k$ and censoring $C_{k+1}$ are physically randomized at each $k$
possibly depending on treatment and covariate history $(\overline{L}_k,%
\overline{A}_{k-1})$. This condition is not, however, guaranteed to hold in
an observational study, or in a randomized trial with censoring, and cannot be
empirically examined. Exchangeability is sometimes referred to as the
assumption of ``no unmeasured confounding'' and $\overline{L}_k$ the
``measured confounder history through $k$''.

\item Positivity: 
\begin{align}
& f_{\overline{A}_{k-1},\overline{L}_{k},C_{k}}(\overline{a}_{k-1}^{g},%
\overline{l}_{k},0)\neq 0\implies  \notag \\
& \Pr[C_{k+1}=0|\overline{L}_k=\overline{l%
}_{k},\overline{A}_{k}=\overline{a}_{k}^{g},C_k=0)\times\notag\\
& f^{obs}(a_k^{g}|\overline{a}_{k-1}^{g},%
\overline{l}_{k},\overline{C}_k=0)>0\;w.p.1.  \label{positivity}
\end{align}%
where $f^{obs}(a_k^{g}|\overline{a}_{k-1}^{g},%
\overline{l}_{k},\overline{C}_k=0)\equiv f_{A_k|\overline{A}_{k-1},\overline{L}_k,\overline{C}_k}(a_k^{g}|\overline{a}_{k-1}^{g},%
\overline{l}_{k},\overline{0})$
\item Consistency: If $\overline{A}_{K}=\overline{a}_{K}^{g}$ and $\overline{C}_{K+1}=0$ then $%
Y=Y^{g}$ and $\overline{L}_{K}=%
\overline {L}_{K}^{g}$.

\end{enumerate}

 Given the three conditions above hold
for all $g\in\mathcal{G}$  such that positivity holds when we replace the observed treatment density $f^{obs}(a_k^{g}|\overline{a}_{k-1}^{g},%
\overline{l}_{k},\overline{C}_k=0)$ with the intervention density $f^{int}(a_k^{g}|\overline{a}_{k-1}^{g},%
\overline{l}_{k},\overline{C}_k=0)$ , then  $\sum_{g\in\mathcal{G}} w(g)\mbox{E}[%
Y^{g}]$  is equivalent to the g-formula characterized by $f^{int}(a_{k}|\overline{l}_{k},\overline{a}%
_{k-1},\overline{C}_{k}=0)$ \citep{robinsfail}%
:
\begin{align}
\sum_{\overline{a}_{K}} \sum_{\overline{l}_{K}}&\mbox{E} [Y|%
\overline{L}_{K}=\overline{l}_{K},\overline{A}_{K}=\overline{a}_{K},%
\overline{C}_{K}=0]\times\nonumber\\
&\prod_{j=0}^{K}\{f(l_{j}|\overline{l}_{j-1},\overline{a}_{j-1},\overline{C}_{j}=0)
f^{int}(a_{j}|\overline{l}_{j},\overline{a}_{j-1},\overline{C}_{j}=0)\}
\label{gformrand}
\end{align}%
where $f(l_{k}|\overline{l}_{k-1},\overline{a}_{k-1},\overline{C}_{k}=0)$
is the observed joint density of the
confounders at $k$ conditional on treatment and covariate history, and remaining uncensored, through $k$.  We use a summation symbol in (\ref{gformrand}) and elsewhere for notational simplicity. However, in general, when $(\overline{A}_k,\overline{L}_k)$ contains any continuous components, sums would be replaced with integrals. 

When interest is in a survival outcome, expression (\ref{gformrand}) is the g-formula for risk of the event of interest by $K+1$ under intervention $f^{int}(a_{k}|\overline{l}_{k},\overline{a}%
_{k-1},\overline{C}_{k}=0)$ which, pulling the implicit prior survival indicator $Y_k$ out of $L_k$, can be more explicitly written as:
\begin{align}
\sum_{\overline{a}_{K}}\sum_{\overline{l}_{K}}\sum_{k=0}^{K}&\Pr [Y_{k+1}=1|\overline{L}_{k}=%
\overline{l}_{k},\overline{A}_{k}=\overline{a}_{k},\overline{C}_{k+1}=\overline{Y}%
_{k}=0]\times\nonumber\\
&\prod_{j=0}^{k}\{f(l_{j}|\overline{l}_{j-1},\overline{a}_{j-1},\overline{C}_{j}=0)
f^{int}(a_{j}|\overline{l}_{j},\overline{a}_{j-1},\overline{C}_{j}=0)\nonumber\\
&\Pr [Y_{j}=0|\overline{L}_{j-1}=\overline{l}_{j-1},%
\overline{A}_{j-1}=\overline{a}_{j-1},\overline{C}_{j}=\overline{Y}%
_{j-1}=0]\} \label{gform1}
\end{align}

As above, when the survival outcome is subject to competing events, we may choose to treat competing events ($D_k$) as an implicit component of $C_{k}$ or $L_k$, $k=0,\ldots,K+1$.  Under the former, expression (\ref{gform1}) will not depend on the distribution of competing events.  Under the latter, it will depend on this distribution; in this case, we can more explicitly write expression (\ref{gform1}) as
\begin{align}
\sum_{\overline{a}_{K}}\sum_{\overline{l}_{K}}\sum_{k=0}^{K}&\Pr [Y_{k+1}=1|\overline{L}_{k}=%
\overline{l}_{k},\overline{A}_{k}=\overline{a}_{k},\overline{C}_{k+1}=\overline{D}_{k+1}=\overline{Y}%
_{k}=0]\times\nonumber\\
&\prod_{j=0}^{k}\{f(l_{j}|\overline{l}_{j-1},\overline{a}_{j-1},\overline{C}_{j}=\overline{D}_j=\overline{Y}_j=0)
f^{int}(a_{j}|\overline{l}_{j},\overline{a}_{j-1},\overline{C}_{j}=\overline{Y}_j=0)\nonumber\\
&\Pr [Y_{j}=0|\overline{L}_{j-1}=\overline{l}_{j-1},%
\overline{A}_{j-1}=\overline{a}_{j-1},\overline{C}_{j}=\overline{D}_j=\overline{Y}%
_{j-1}=0]\nonumber\\
&\Pr [D_{j+1}=0|\overline{L}_{j}=\overline{l}_{j},%
\overline{A}_{j}=\overline{a}_{j},\overline{C}_{j+1}=\overline{D}_j=\overline{Y}%
_{j}=0]\} \label{gform2}
\end{align}
See \cite{crpaper} for details.  

Conditions required for identification based on only observed variables of the mean/risk of $Y$ under an alternative intervention characterized by $f^{d}(a_{k}|\overline{a}^{*}_k,\overline{l}_{k},\overline{a}%
_{k-1},\overline{C}_k=0)$, which additionally depends on the history of the natural value of treatment, are generally stronger than those defined above.  We refer the reader elsewhere for details of these conditions \citep{swigs,youngthreshold}.  Provided these stronger conditions hold, the mean of $Y$ under an intervention characterized by $f^{d}(a_{k}|\overline{a}^{*}_k,\overline{l}_{k},\overline{a}%
_{k-1},\overline{C}_k=0)$ is equivalent to the extended g-formula of \cite{WHOchap}
\begin{align}
\sum_{\overline{a}^{*}_{K}}\sum_{\overline{a}_{K}} \sum_{\overline{l}_{K}}&\mbox{E} [Y|%
\overline{L}_{K}=\overline{l}_{K},\overline{A}_{K}=\overline{a}_{K},%
\overline{C}_{K}=0]\times\nonumber\\
&\prod_{j=0}^{K}\{f(l_{j}|\overline{l}_{j-1},\overline{a}_{j-1},\overline{C}_{j}=0)
f^{d}(a_{j}|a^{*}_j,\overline{l}_{j},\overline{a}_{j-1},\overline{C}_{j}=0)\nonumber\\
&f(a^{*}_{j}|\overline{l}_{j},\overline{a}_{j-1},\overline{C}_{j}=0)\}
\label{gformrandastar}
\end{align}%
where $f(a^{*}_{j}|\overline{l}_{j},\overline{a}_{j-1},\overline{C}_{j}=0)$ is the observed treatment density conditional on the measured past evaluated at some possibly realized values $(\overline{A}_j,\overline{L}_j)=(a^{*}_{j},\overline{a}_{j-1},\overline{l}_j)$.  

In motivating the general estimation algorithm described in the next section, it is useful to consider a generic version of the g-formula for the mean/risk of the outcome characterized by either an intervention density $f^{d}(a_{k}|\overline{a}^{*}_k,\overline{l}_{k},\overline{a}%
_{k-1},\overline{C}_k=0)$ which depends on the natural value of treatment at $k$ or $f^{int}(a_{k}|\overline{l}_{k},\overline{a}%
_{k-1},\overline{C}_k=0)$  which depends, at most, on the measured past treatment and confounder history.  For $Z_k=(L_k,A_k)$ and realization $z_k=(l_k,a^{*}_k)$, $k=0,\ldots,K$ we have : 
\begin{align}
\sum_{\overline{a}_{K}} \sum_{\overline{z}_{K}}&\mbox{E} [Y|%
\overline{L}_{K}=\overline{l}_{K},\overline{A}_{K}=\overline{a}_{K},%
\overline{C}_{K}=0]\times\nonumber\\
&\prod_{j=0}^{K}\{f(z_{j}|\overline{l}_{j-1},\overline{a}_{j-1},\overline{C}_{j}=0)h^{user}(\overline{a}_j,a^{*}_j,\overline{l}_j)\}
\label{gformgeneric}
\end{align}%
is algebraically equivalent to (\ref{gformrandastar}) when  $h^{user}(\overline{a}_k,a^{*}_k,\overline{l}_k)$ is chosen as some $f^{d}(a_{k}|\overline{a}^{*}_k,\overline{l}_{k},\overline{a}%
_{k-1},\overline{C}_k=0)$ and algebraically equivalent to (\ref{gformrand}) when $h^{user}(\overline{a}_k,a^{*}_k,\overline{l}_k)$ is alternatively chosen as some $f^{int}(a_{k}|\overline{l}_{k},\overline{a}%
_{k-1},\overline{C}_k=0)$ as, in this latter case, the expression does not depend on the observed treatment density $f(a^{*}_{k}|\overline{l}_{k},\overline{a}_{jk1},\overline{C}_{k}=0)$ \citep{youngthreshold}.  

Similarly, for survival outcomes, we can expressly write the g-formula for risk by $K+1$ under generic treatment interventions $h^{user}(\overline{a}_j,a^{*}_j,\overline{l}_j)$ 
\begin{align}
\sum_{\overline{a}_{K}}\sum_{\overline{z}_{K}}\sum_{k=0}^{K}&\Pr [Y_{k+1}=1|\overline{L}_{k}=%
\overline{l}_{k},\overline{A}_{k}=\overline{a}_{k},\overline{C}_{k+1}=\overline{D}_{k+1}=\overline{Y}%
_{k}=0]\times\nonumber\\
&\prod_{j=0}^{k}\{f(z_{j}|\overline{l}_{j-1},\overline{a}_{j-1},\overline{C}_{j}=\overline{D}_j=0=\overline{Y}_j=0)h^{user}(\overline{a}_j,a^{*}_j,\overline{l}_j)\nonumber\\
&\Pr [Y_{j}=0|\overline{L}_{j-1}=\overline{l}_{j-1},%
\overline{A}_{j-1}=\overline{a}_{j-1},\overline{C}_{j}=\overline{D}_j=\overline{Y}%
_{j-1}=0]\nonumber\\
&\Pr [D_{j+1}=0|\overline{L}_{j}=\overline{l}_{j},%
\overline{A}_{j}=\overline{a}_{j},\overline{C}_{j+1}=\overline{D}_j=\overline{Y}%
_{j}=0]\} \label{gform2generic}
\end{align}
when the outcome is subject to competing events and competing events are not treated as censoring events.  Following arguments above and in \cite{crpaper}, replacing $ \Pr [D_{j+1}=0|\overline{L}_{j}=\overline{l}_{j},%
\overline{A}_{j}=\overline{a}_{j},\overline{C}_{j+1}=\overline{D}_j=\overline{Y}%
_{j}=0] $ with 1 for all $j$ and $(\overline{a}_j,\overline{l}_j)$ would give the g-formula for risk by $K+1$ under an intervention $h^{user}(\overline{a}_j,a^{*}_j,\overline{l}_j)$  when competing events are treated as censoring events and, thus, under an intervention that eliminates competing events along with other possible forms of censoring.

\section{Estimation algorithm}\label{algorithm}

\subsection{Survival outcomes}\label{survalgorithm}
The following describes the general computational algorithm for estimating the g-formula for risk of the outcome (the event of interest by $K+1$) under a user-specified intervention $h^{user}(\overline{a}_k,a^{*}_k,\overline{l}_k)$.  This corresponds to (i) expression (\ref{gform2generic}) when competing events are not treated as censoring events; (ii) a restricted version of (\ref{gform2generic}) that  replaces $\Pr [D_{j+1}=0|\overline{L}_{j}=\overline{l}_{j},%
\overline{A}_{j}=\overline{a}_{j},\overline{C}_{j+1}=\overline{D}_j=\overline{Y}%
_{j}=0] $ with 1 when competing events are treated as censoring events for all $j$ (i.e. an implicit component of $C_{j+1}$) and $(\overline{a}_j,\overline{l}_j)$; and (iii) a restricted version of (\ref{gform2generic}) with $D_k\equiv0$ for all $k$ when the outcome is not subject to competing events.
 
Let $Z_k=(L_k,A_k)$ such that for all $k = 0, \dots, K$,
\begin{equation*}
f(Z_k|\overbar{L}_{k-1}, \overbar{A}_{k-1}, \overbar{C}_k =\overline{D}_k= \overbar{Y}_{k}=0) = \prod_{j=1}^{p} f(Z_{j,k}|Z_{j-1,k},\ldots,Z_{1,k}, \overline{L}_{k-1},\overbar{A}_{k-1}, \overbar{C}_k =\overline{D}_k= \overbar{Y}_{k}=0)
\end{equation*}
where $Z_{j,k}$ is the $j^{th}$ component of the vector $Z_k$, $j=1,\ldots,p$ for a, generally, arbitrary permutation of these $p$ components.  Note that certain permutations will be preferred when \textsl{a priori} knowledge of the distributions of certain components of $Z_k$ conditional on values of other components is known (see Section \ref{deterministic}).

For a user-chosen value of $K$, treatment rule $h^{user}(\overline{a}_k,a^{*}_k,\overline{l}_k)$ and permutation of $Z_k$, $k=0, \dots, K$,  we apply the following algorithm to a subject-interval input data set constructed according to the instructions of Section \ref{datastructure}:

\begin{enumerate}
	\item Using all subject interval records:
	
	\begin{enumerate}
		\item If $k > 0$, estimate the conditional densities $f(Z_{j,k}|Z_{j-1,k},\ldots,Z_{1,k}, \overline{L}_{k-1},\overbar{A}_{k-1}, \overbar{C}_k =\overline{D}_k= \overbar{Y}_{k}=0)$ under a distributional assumption on $Z_{j,k}$ given ``history'' \\
		$(Z_{j-1, k}, \dots, Z_{1, k}, \overbar{L}_{k-1}, \overbar{A}_{k-1})$, $j = 1, \dots, p$.   This distribution will be used to simulate a value of $Z_{j, k}$ in  Step 2 at each $k>0$.  For example, dichotomous $Z_{j, k}$ will, by default, be generated from a Bernoulli distribution with mean $\mu(Z_{j-1, k}, \dots, Z_{1, k}, \overbar{L}_{k-1}, \overbar{A}_{k-1}) = \Pr[ Z_{j, k} = 1|Z_{j-1, k}, \dots, Z_{1, k}, \overbar{L}_{k-1}, \overbar{A}_{k-1}, C_k = D_k=Y_k = 0]$. For  $Z_{j,k}$ with many levels, the user can select among several pre-specified distributions (e.g. Normal) or can rely on a user-supplied distribution.   The user has the option to estimate the conditional mean of $Z_{j,k}$  $\mu(Z_{j-1, k}, \dots, Z_{1, k}, \overbar{L}_{k-1}, \overbar{A}_{k-1})$ for a specified function of the ``history'' $(Z_{j-1, k}, \dots, Z_{1, k}, \overbar{L}_{k-1}, \overbar{A}_{k-1})$ using pre-specified fit options (e.g. generalized linear models with specified link function) or other user-supplied functions (e.g. generalized additive models, classification and regression trees) that produce a fitted object based on one line of data.  Syntax for these specifications is described in Section \ref{covdist} via specification of the parameters \verb|covtypes| and \verb|covparams| .  These conditional mean models are pooled over time $k$ and thus the user must specify the function of time $k$, if any, on which they will depend (see Sections \ref{covdist} and \ref{discrete_time}).  When the variance of $Z_{j,k}$ is not a function of the mean under the distributional assumption (e.g. when $Z_{j,k}$ is assumed Normal), the variance is automatically estimated by the model residual mean squared error.  
				
		\item Estimate the conditional probability (discrete hazard) of the event of interest at each time $k+1$ $k=0,\ldots,K$ conditional on treatment and covariate history and  surviving and remaining uncensored to the previous time, $p_k(\overline{l}_k,\overline{a}_k)\equiv\Pr[Y_{k+1} = 1|\overbar{L}_{k} = \overbar{l}_{k}, \overbar{A}_{k} = \overbar{a}_{k}, \overbar{C}_{k+1} = \overline{D}_{k+1}=0, \overbar{Y}_{k} = 0]$.   These estimates are automatically based on fitting a pooled over time logistic regression model with $Y_{k+1}$ the dependent variable where the user specifies the function of ``history''  $(\overline{L}_k,\overline{A}_k)$ included as independent variables in the fit. See Section \ref{outcome}.  		
		\item If the event of interest is subject to competing events and competing events are not defined as censoring events: Estimate the conditional probability (discrete hazard) of the competing event at each time $k$ conditional on treatment and covariate history and surviving and remaining uncensored to the previous time $q_k(\overline{l}_k,\overline{a}_k)\equiv\Pr[D_{k+1} = 1|\overbar{L}_{k} = \overbar{l}_{k}, \overbar{A}_{k} = \overbar{a}_{k}, \overbar{C}_{k+1} = \overline{D}_{k}=0, \overbar{Y}_{k} = 0]$.  These estimates are automatically based on fitting a pooled over time logistic regression model with $D_{k+1}$ the dependent variable where the user specifies the function of ``history''  $(\overline{L}_k,\overline{A}_k)$ and time included as independent variables in the fit. See Section \ref{outcome}.  
	\end{enumerate}

	\item Select a value $s\geq 10,000$, possibly such that $s=n$.  If $s\neq n$ resample the original $n$ ids $s$ times with replacement and create a new data set where each possibly resampled record has a unique id $v=1,\ldots,s$.  Using the baseline covariate data from the original $v=1,\ldots,n$ observations at baseline (if $s=n$) or the resampled observations (if $s\neq n$), for $k = 0, \dots, K$ and $v = 1, \dots, s$, do the following:
	
	\begin{enumerate}
		\item If $k = 0$, set $z_{0, v}$ to the observed values of $Z_{0}$ for
		id $v$. Otherwise, if $k > 0$, iteratively draw $z_{j,k, v}$ $j=1,\ldots,p$ from estimates of the conditional densities $f(Z_{j,k}|Z_{j-1,k},\ldots,Z_{1,k}, \overline{L}_{k-1},\overbar{A}_{k-1}, \overbar{C}_k =\overline{D}_k= \overbar{Y}_{k}=0)$ (based on specifications from step 1.a) evaluated at $Z_{j,k}=z_{j,kv}$, $(Z_{1,k},\ldots,Z_{j-1,k},\overline{L}_{k-1}) = (z_{1,k,v},\ldots,z_{j-1,k,v},\overbar{l}_{k-1,v})$, the previously drawn covariates
		and $\overline{A}_{k-1}=\overline{a}^{user}_{k-1,v}$, the previously assigned treatment through $k-1$
		under the user-chosen strategy. 
		
		\item Denote the observed (at $k=0$) or drawn (at $k>0$) value(s) of treatment(s) in the vector $z_{k, v}$ as $a_{k, v}^{*}$.  Assign $a^{user}_{k,v}$ according to the user specified rule $h^{user}(\overline{a}^{user}_{k,v},a^{*}_k,\overline{l}_{k,v})$ which may or may not depend on $a^{*}_{k,v}$.  For example, suppose $h^{user}$ is defined as the static deterministic strategy ``always set treatment to the value 1.''  In this case,  $a^{user}_{k, v}$ is always set to 1.  Another example is ``if the natural value of treatment at $k$ is less than 30, set treatment to 30.  Otherwise, do not intervene.'' In this case, the observed/drawn value of $a^{*}_{k,v}$ is examined.  If $a^{*}_{k,v}<30$ then  $a^{user}_{k, v}$ is set to 30.  Otherwise, $a^{user}_{k, v}$ is set to $a^{*}_{k,v}$.  A third example is ``no intervention on $A_k$'' or the so-called \textsl{natural course} intervention.  In this case, the intervention density $h^{user}$ is defined as the observed treatment density and $a^{user}_{k, v}$ is always set to $a^{*}_{k,v}$.  See Sections \ref{strategies} and \ref{customint} for required syntax for specifying the treatment rules.
				
		\item Estimate $p_k(\overline{l}_{k,v},\overline{a}^{user}_{k,v})$ based on specifications from step 1.b. Denote this estimate $\hat{p}(\overline{l}_{k,v},\overline{a}^{user}_{k,v})$.
		
		\item If the event of interest is subject to competing events and competing events are not defined as censoring events: Estimate $q_k(\overline{l}_{k,v},\overline{a}^{user}_{k,v})$ based on specifications from step 1.c.  Denote this estimate $\hat{q}(\overline{l}_{k,v},\overline{a}^{user}_{k,v})$.
		
	\end{enumerate}
	
	\item Compute the intervention risk estimate as
	
\begin{equation}
	\frac{1}{n}\sum_{v=1}^{n}\sum_{k=0}^{K}\hat{p}(\overbar{a}^{user}_{k,v},\overbar{l}_{k,v}) \prod_{j=0}^{k}\{1-\hat{p}(\overbar{a}^{user}_{j,v},\overbar{l}_{j,v})\}\{1-\hat{q}(\overbar{a}^{user}_{j+1,v},\overbar{l}_{j+1,v})\}\label{estimatorsurv}
	\end{equation}
	In the special case where either the event of interest is not subject to competing events or competing events are treated as censoring events $\hat{q}(\overline{l}_{j+1,v},\overline{a}^{user}_{j+1,v})$ is set to 0 for all $j$ in (\ref{estimatorsurv}).
	
	\end{enumerate}
	Steps 2 and 3 above can be repeated for multiple user-defined treatment rules $h^{user}$ and risk ratio/risk difference estimates constructed based on a specified reference intervention.  The natural course intervention is always automatically implemented and is the default reference unless otherwise specified by the user.  Ninety-five percent confidence intervals are computed based on repeating the entire algorithm (Steps 1-3) in $B$ bootstrap samples based on percentiles of the bootstrap distribution for user specified $B$.  The user can also opt to additionally output estimates of the risk (risk ratios/risk differences) by all times $t+1$, $t\leq K$ in addition to the risk by $K+1$.  
	
	In addition to the parametric g-formula estimate of the natural course risk, a nonparametric estimate will also be computed automatically and reported in the output.  When censoring events are present in the data, this is a completely unadjusted estimate of the natural course risk (i.e. it relies on marginal exchangeability for censoring)\citep{pgformyoung}. When competing risks are absent or treated as censoring events, this is computed as the complement of a product-limit survival estimator \citep{km} in the observed data.  When competing risks are present and not treated as censoring events, this is computed using an Aalen Johansen estimator \citep{aalenjohansen}.  See \cite{crpaper} and \cite{documentation} for additional details.  
	
\subsection{Fixed end of follow-up outcomes}\label{eofalgorithm}
The algorithm for estimating the g-formula for the mean of an outcome at the end of follow-up $K+1$ under a user-specified intervention $h^{user}(\overline{a}_k,a^{*}_k,\overline{l}_k)$ is nearly identical to that for the risk in the case of survival outcomes described in the previous section but relies on a slightly different input data structure (see Section \ref{datastructure}).  In the modified algorithm \\

Step 1.b. is replaced by
\begin{itemize}
\item Modified Step 1.b.: Using only records on line $K+1$, estimate the mean of the outcome at $K+1$ conditional on treatment and covariate history and remaining uncensored to the previous time, $\mu(\overline{l}_K,\overline{a}_K)\equiv \mbox{E}[Y|\overbar{L}_{K} = \overbar{l}_{K}, \overbar{A}_{k} = \overbar{a}_{K}, \overbar{C}_{K+1} = 0]$.   This estimate is automatically based on fitting a linear regression model with $Y$ the dependent variable where the user specifies the function of ``history''  $(\overline{L}_K,\overline{A}_K)$.  Because this fit will only use records on line $K+1$ of the data it will not be a function of time.  See Section \ref{outcome}.
\end{itemize}
Step 2.c. is replaced by:
\begin{itemize}
\item Modified Step 2.c.: For $k=K$ only, estimate $\mu(\overline{l}_K,\overline{a}_K)$ based on specifications from step 1.b.   Denote this estimate $\hat{\mu}(\overline{l}_{K,v},\overline{a}^{user}_{K,v})$.
\end{itemize}
Step 3 is replaced by 
\begin{itemize}
\item Modified Step 3: Compute the intervention mean estimate as
\begin{equation}
	\frac{1}{n}\sum_{v=1}^{n}\hat{\mu}(\overbar{a}^{user}_{K,v},\overbar{l}_{K,v}) \label{estimatormean}
	\end{equation}
\end{itemize}
As discussed above, in this case, if a subject dies prior to $K+1$, this must be treated as a censoring event ($D_{k+1}$ must be considered a component of the vector $C_{k+1}$, $k=0,\ldots,K$) and steps 1.c and 2.d are not implemented.

\section[Using the gformula package]{Using the \pkg{gfoRmula} package}\label{package}

\subsection{Specifying the outcome type}

The \pkg{gfoRmula} package gives the user access to three different \verb|gformula| function variants, each intended for a different outcome type. \verb|gformula_survival| should be used for survival outcomes (e.g. an indicator of all-cause mortality by $K+1$); \verb|gformula_continuous_eof| should be used for fixed continuous end of follow-up outcomes (e.g. blood pressure at $K+1$); and \verb|gformula_binary_eof| should be used for fixed binary end of follow-up outcomes (e.g. indicator of high cognitive score at $K+1$).


\subsection{Required structure of the input dataset}\label{datastructure}

The input dataset to the \verb|gformula| function must be an \proglang{R} \verb|data.table|. For all outcome types, each row of the input dataset should contain one record for each time $k$ (which must be a numeric variable in \proglang{R}, and the name of which is specified by the parameter \verb|time_name|) for each subject present at baseline (specified by the parameter \verb|id|).  The index $k$ must start at 0 and increase in increments of 1. Each column of the data set will index a time-varying covariate $Z_{j,k}$, $j=1,\ldots,p$. A subject's baseline confounders (the time-fixed components of $L_0$) should be repeated on each line $k$ for that subject in the data set. A subject who is censored in interval $k+1$ $(C_k+1=1)$ will have only $k+1$ records with time index $k$ on the last line.  Other requirements are dependent on the outcome type which we describe below.  Each subject can have a maximum of $K+1$ records with $K+1$, the end of follow-up of interest.  

\subsubsection{Survival outcomes}

In this case, the data set should additionally contain a column $Y_{k+1}$ indicating, on line $k$ whether the event of interest occurred by the next interval.   A subject who first has the event of interest in interval $k+1$ will have only $k+1$ records with time index $k$ on the last line.  If $C_{k+1}=1$ then the user has the choice to code $Y_{k+1}$ as either \verb|NA| or 0. Records coded as \verb|NA| will be excluded when fitting the discrete hazard model in step 1.b of the algorithm and in nonparametric estimation of the natural course risk (see Section \ref{algorithm}) while records coded as 0 will be included in computing these estimates. This choice impacts how ``risk sets'' at $k$ are defined.  It will make no difference to the estimates when the intervals are made small enough such that there are no failures in intervals where there are also censoring events.  Otherwise, this choice may impact effect estimates to some degree.  Note that by (i) the fact that the algorithms of the previous section do not rely on an estimate of the distribution of censoring and (ii) the required structure of the input data set such that subjects have no more records after they are censored, the input data set does not actually require any variables defining censoring indicators.  

If the event of interest is subject to a competing event $D_{k+1}\ne 0$ for some subjects and some $k$ \textsl{and} the user chooses to treat competing events as censoring events then $D_{k+1}$ should be treated like any component of $C_{k+1}$ and coding requirements outlined above must be applied.  In turn, in this case, the input data set does not require a variable for $D_{k+1}$.   If, alternatively, competing events are not treated as censoring events, then the user \textsl{must} include in the input data set a time-varying indicator of the competing event $D_{k+1}$.  If $D_{k+1}=1$ for a given subject, then that subject will only have $k+1$ lines in the data with time index $k$ on the last line and, on that line, $Y_{k+1}$ must be coded \verb|NA|.   See the data sets \verb|basicdata_nocomp| and \verb|basicdata| included with the package for examples of input data sets for survival outcomes without and with competing events, respectively.  The user can specify the parameter \verb|time_points| to be a follow-up interval $k$ in order to compute the risk estimates through $k<K+1$ (with $K+1$ the max number of records for an individual in the data set).  The default value of \verb|time_points| is $K+1$.    

\subsubsection{End of follow-up outcomes}

In this case, the data set should additionally contain a column $Y$ with the value of the outcome in interval $K+1$.  Only the value of this variable on line $k=K$ is used in the algorithm and it does not matter what values are coded on lines $k<K$.  These may be left blank or given any value.  As discussed above, death or other truncation events, if present, must be treated as censoring events in this case.  For subjects who are first censored in interval $K+1$, these subjects will have $K+1$ records in the data and $Y$ on line $k=K$ must be coded as \verb|NA|.  For subjects censored prior to interval $K+1$, as above, the value of $Y$ will be ignored in the estimation algorithm. See the data set \verb|continuous_eofdata| included with the package for an example of an input data set for an end of follow-up outcome.  Unlike for survival outcomes, the default value of \verb|time_points| ($K+1$) cannot be changed for end of follow-up outcomes.

\subsection{Specifying the input data set, individual identifier, time index, outcome and competing events}

The input data set is specified by the argument \verb|obs_data|, the column of this data set specifying the individual identifier by the argument \verb|id|, the column of this data set specifying the time index by the argument \verb|time_name|, the column of this data set specifying the outcome by the argument \verb|outcome_name| and, if present and not defined as a censoring event, the column of this data set specifying a competing event by the argument \verb|compevent_name|.  Sample syntax is given by:

\begin{verbatim}
gformula_survival(..., obs_data = obs_data,
                  id = 'ID',
                  time_name = 't0',
                  outcome_name = 'Y',
                  compevent_name = 'D')
\end{verbatim}
The argument \verb|compevent_name| should only be specified for survival outcomes and when the user chooses not to treat competing events as censoring events. Note that the input data set must be a \verb|data.table| object. 

\subsection{Specifying time-varying and baseline covariates}\label{covariates}

The time-varying covariates $Z_k$ are defined by the vector argument \verb|covnames|.  Baseline confounders (the time-fixed components of $L_0$) are defined by the vector argument \verb|basecovs|.  Sample syntax is given by:
\begin{verbatim}
gformula_survival(..., covnames = c('Z1', 'Z2', 'Z3'),
                       basecovs = c('race','sex')		)
\end{verbatim}
When the user chooses to allow outcome and/or covariate distributions to be certain discrete functions of time, an additional argument must be added to \verb|covnames| (see Section \ref{discrete_time}).

\subsection{Generating covariate histories}\label{histories}
Recall that, in step 1.a. of the estimation algorithm (Section \ref{algorithm}), the user must specify a function of ``history'' when estimating the conditional distribution of each $Z_{j,k}$ conditional on  ``history'' $(Z_{j-1, k}, \dots, Z_{1, k}, \overbar{Z}_{k-1})$ for each $j=1,\ldots,p$.  For survival outcomes, the user must specify a function of ``history'' when estimating the observed conditional hazards of the event of interest at each $k+1$ in step 1.b. (and of the competing event in step 1.c. if competing events are not treated as censoring events) conditional on ``history'' $\overline{Z}_k$.  For end of follow-up outcomes, the user must specify a function of ``history'' when estimating the mean of the outcome at $Y_{K+1}$ conditional on ``history'' $\overline{Z}_K$ only.  

The arguments \verb|histvars| and \verb|histories| must be used to specify any desired functions of history that will be used for estimation in step 1 (and then, in turn, in the simulation in step 2) that cannot be defined within a model statement in \proglang{R} .  For example, consider a time-varying covariate $Z_{4,k}$ named \verb|Z4| in the input data set and suppose the user assumes that the distributions of ``future'' covariates $Z_{j,k}$, $j>4$ depend on the history of $Z_{4,k}$ (or $\overline{Z}_{4,k}$) through the cumulative average of this variable through each $k$; i.e. $\frac{1}{k}\sum_{t=0}^{k} Z_{4,t}$ for $k>0$ and $Z_{4,0}$ otherwise.  Then the user must use \verb|histvars| and \verb|histories| to create in the data set \verb|obs_data| this additional time-varying covariate containing the cumulative average of $Z_{4,k}$ at each $k$. Alternatively, if the user made the assumption that the distributions of these ``future'' covariates/outcomes only depend on $Z_{4,k}$ through the current value at $k$ \textsl{or} a transformation of the current value (i.e. square root, restricted cubic spline, quadratic or cubic function) which can be made within an \proglang{R} model statement, then no additional covariates need to be created as these transformations can be made within a model statement (see Section \ref{covdist}).

Desired functions of history that cannot be created within a model statement in \proglang{R} must be created using \verb|histvars| and \verb|histories| to ensure that estimation decisions in step 1 are carried forward correctly through the simulation in step 2.   Pre-coded functions of history for a covariate named \verb|Zj|  in the input data set include:
\begin{itemize}
\item \verb|lagged|: this adds a variable to the specified input data set named \verb|lagi_Zj|, containing the $i$th lag of \verb|Zj| relative to the current time $k$ (i.e. its value at $k-1$ for $k\geq i$) for all $i=1,\ldots,r$ with $r$ the desired number of lags.  \verb|lagi_Zj| is set to 0 on lines with $k<i$  
\item \verb|cumavg|:  this adds a variable to the specified input data set named \verb|cumavg_Zj|, which contains the cumulative average of \verb|Zj| up until the current time $k>0$.  It is set to \verb|Zj| at $k=0$.
\item \verb|lagavg|: this adds a variable to the specified input data set named \verb|lag_cumavgi_Zj| which contains the $i$th lag of the cumulative average of \verb|Zj| relative to the current time $k$, $i=1,\ldots,r$. \verb|lag_cumavgi_Zj| is set to 0 on lines with $k<i$.   
\end{itemize}
Note the desired number of lags $r$ is specified using the argument \verb|covmodels| (see Section \ref{covdist}).

The package will apply the function of history listed in the $q$th element of the vector argument \verb|histories| to all variables listed in the $q$th element of the list argument \verb|histvars|. Therefore, the length of \verb|histvars| and \verb|histories| need to match.  Sample syntax that would add lagged, cumulative average and lagged cumulative average functions of the covariates \verb|Z1|, \verb|Z2| and \verb|Z3| as new variables to the input data set \verb|dataname| for use in estimation of conditional distributions/hazards/means (see Sections \ref{covdist}-\ref{outcome}) is as follows:
\begin{verbatim}
gformula_survival(..., obs_data = dataname, covnames = c('Z1', 'Z2', 'Z3'),
                  basecovs = c('race', 'sex')	,
                  histories = c(lagged, cumavg, lagavg),
                  histvars = list(c('Z1', 'Z2', 'Z3'), c('Z1', 'Z2', 'Z3'), 
                                  c('Z1', 'Z2', 'Z3')))
\end{verbatim}

\subsection{Specifying covariate distributions}\label{covdist}

The vector argument \verb|covtypes| is used to specify the distributions of each time-varying covariate $Z_{j,k}$ conditional on a function of history $(Z_{j-1, k}, \dots, Z_{1, k}, \overbar{Z}_{k-1})$, $j=1,\ldots,p$ and $k=0,\ldots,K$, how parameters of those distributions will be estimated and, in turn, how covariate values in step 2 of the algorithm will be simulated.  The \pkg{gfoRmula} package supplies a number of pre-programmed options for input to the argument \verb|covtypes| which are described below: \verb|'binary'|, \verb|'normal'|, \verb|'categorical'|, \verb|'bounded normal'|, \verb|'zero-inflated normal'|, \verb|'truncated normal'|, \verb|'absorbing'| and   \verb|'categorical time'|.  Each of the elements of \verb|covtypes| requires its own set of specific sub-parameters, which are contained within the list argument \verb|covparams|.  Each sub-parameter vector within this list must be the same length as \verb|covnames| and \verb|covtypes|. 

We now give a description of each of the pre-programmed options for input to \verb|covtypes| along with required and optional sub-parameters:
\begin{itemize}
	\item \verb|'binary'|: The mean of the covariate conditional on history (see Section \ref{histories}) is estimated via the estimated coefficients of a generalized linear model (GLM) (using \verb|glm| in \proglang{R}), where the family of the GLM is \verb|binomial|. The covariate values are simulated at each time $k$ in step 2 of the estimation algorithm (Section \ref{algorithm}) given the previously simulated history under intervention by sampling from a Bernoulli distribution with parameter this estimated conditional mean. Assuming the covariate in question is the $j$th entry in \verb|covnames|, then the sub-parameter \verb|covmodels|, which is a vector of length \verb|covnames|, must contain as its $j$th entry a model statement for estimating the mean of the covariate in the $j$th entry of \verb|covnames| as a function of history and possibly \verb|time_name|. Optionally, the user can also specify the $j$th entry of \verb|covlink|, which is a vector containing the link function(s) for the GLM specified in the $j$th entry of \verb|covmodels|. The default value of \verb|covlink| for the \verb|'binary'| covtype is \verb|logit|. Note that binary covariates in the input data set must be of class numeric and coded as 0 or 1.
	\item \verb|'normal'|: The mean of the covariate conditional on history is estimated via the estimated coefficients of a GLM where the family is \verb|gaussian|. The covariate values are simulated at each time $k$ in step 2 given the previously simulated history under intervention by sampling from a normal distribution with mean this estimated conditional mean and variance the residual mean squared error from the GLM fit (see Section \ref{algorithm}). Values generated outside the observed range for that covariate over all times $k$ are subsequently set to the minimum or maximum of this range. The sub-parameter \verb|covmodels|, defined as in the \verb|binary| case, must be specified, and the sub-parameter \verb|covlink| can optionally also be specified.  The default value of \verb|covlink| for the \verb|'normal'| covtype is \verb|identity|. 
	\item \verb|'categorical'|: The probability that a covariate with at least 3 levels takes a particular level conditional on history, is estimated via the estimated coefficients of a multinomial logistic regression model. Specifically, the \verb|multinom| function in the \pkg{nnet} package is used to fit the model \citep{nnet}. In order to fit the model, the covariate must be made a factor in the input data set (e.g., by using the \verb|as.factor| function). The covariate values are simulated at each time $k$ in step 2 given the previously simulated history under intervention by sampling from a multinoulli distribution with parameters these estimated conditional probabilities. The sub-parameter \verb|covmodels| must be specified.
	\item \verb|'bounded normal'|: The observed covariate values are first standardized to the interval $[0, 1]$, inclusive, by subtracting the minimum value and dividing by the range. Subsequently, the mean of the standardized covariate conditional on history is estimated via the estimated coefficients of a GLM (family \verb|gaussian|). Standardized covariate values are simulated at each time $k$ in step 2 given the previously simulated history under intervention by sampling from a normal distribution with mean this estimated conditional mean and variance the residual mean squared error from the GLM fit. The simulated standardized values are then transformed back to the original scale, and generated values that fall outside the observed range are set to the minimum or maximum of this range. The sub-parameter \verb|covmodels| must be specified.  The sub-parameter \verb|covlink| can optionally be specified. The default value of \verb|covlink| for the \verb|bounded normal| covtype is \verb|identity|.
	\item \verb|'zero-inflated normal'|: For a covariate with support $\ge 0$, the probability that the covariate equals 0 conditional on history is first estimated via the estimated coefficients of a generalized linear model (\verb|glm|) where the family of the GLM is \verb|binomial|. The mean of the covariate values conditional on history and among positive values is then estimated via a GLM where the family is \verb|gaussian|. The simulated covariate values are created by first generating an indicator of whether the covariate value is zero or non-zero from a Bernoulli distribution with the estimated probability of the first model. Covariate values are then generated from a normal distribution with the estimated mean of the second model and multiplied by the generated zero indicator. Non-zero generated covariate values that fall outside the observed range are set to the minimum or maximum of the range \emph{of non-zero observed values} of the covariate. The sub-parameter \verb|covmodels|, defined as before, must be specified, and the sub-parameter \verb|covlink| can optionally also be specified.  The default value of \verb|covlink| for the \verb|'zero-inflated normal'| covtype is \verb|identity|.
	\item \verb|'truncated normal'|: The mean of the covariate is estimated via the estimated coefficients of a truncated normal regression model. The \verb|truncreg| function in the \pkg{truncreg} package is used to fit the model \citep{truncreg}. The simulated covariate values are generated by sampling from a truncated normal distribution with the estimated covariate mean. Generated values that fall outside the observed range are set to the minimum or maximum of the observed range. The sub-parameter \verb|covmodels|, defined as before, must be specified, as well as the sub-parameter \verb|point|. Assuming the covariate is the $j$th entry of \verb|covnames|, \verb|point|, which is a vector of length $p$, must contain as its $j$th entry the truncation point for the covariate, and the sub-parameter \verb|direction|, which is likewise a vector of length $p$, must contain as its $j$th entry the direction(s) of truncation (\verb|'left'| or \verb|'right'|).
	\item \verb|'absorbing'|: This option is similar to \verb|'binary'|, with the same required and optional sub-parameters. However, in this case, the GLM fit is restricted to records where the value of the covariate at $k-1$ (the lagged value) is 0.  Then in the simulation step, once a 1 is first generated, the covariate value at that time and all subsequent times is set to 1.  This option applies to a covariate that, once it switches to 1 at time $k$, it stays 1 at all subsequent times (e.g. an indicator of a disease diagnosis \textsl{by} time $k$.)
	\item \verb|'categorical time'|: See details for this option in Section \ref{discrete_time}.
\end{itemize}

Any sub-parameters required by a specified \verb|covtype| must be specified as a vector element of \verb|covparams| with length equal to the length of \verb|covnames| but some entries in each vector may be \verb|NA|. The following sample syntax illustrates this principle for a dataset with covariates \verb|Z1|, \verb|Z2|, and \verb|Z3|: 
\begin{verbatim}
gformula_survival(..., covnames = c('Z1', 'Z2', 'Z3'), basecovs = c('race', 'sex'),
                       time_name = 't0', 
                       histories = c(lagged),
                       histvars = list(c('Z1', 'Z2', 'Z3')),
                       covtypes = c('categorical', 'truncated normal', 'binary'),
                       covparams = list(covmodels = c(Z1 ~ lag1_Z2 + lag1_Z3 +
                                                      lag1_Z1 + lag2_Z2 + lag2_Z3 + 
                                                      lag2_Z1 
                                                      + race  + sex + t0 + I(t0^2),
                                                      Z2 ~ Z1 + lag1_Z1 + 
                                                      lag1_Z2 + lag1_Z3 + +lag2_Z2 + 
                                                      lag2_Z3 +
                                                      race + sex + t0 + I(t0^2),
                                                      Z3 ~ Z1 + Z2+ lag1_Z1 + 
                                                      lag1_Z2+ lag1_Z3 + lag2_Z3 + race 
                                                      + 
                                                      sex + t0 + I(t0^2)),
                                        covlink = c(NA, NA, 'probit'),
                                        point = c(NA, 0.5, NA),
                                        direction = c(NA, 'left', NA)))
\end{verbatim}

In this example, all three covariates require a model statement, so the \verb|covmodels| sub-parameter contains three non-\verb|NA| entries. However, only the third covariate, \verb|Z3|, requires the subparameter \verb|covlink| so only the third entry of the \verb|covlink| sub-parameter is non-\verb|NA|. Likewise, only the second covariate, \verb|Z2|, requires the subparameters \verb|point| and \verb|direction| so only the second entries of these subparameters  are non-\verb|NA|.
Note in this example, by specifying \verb|histories = c(lagged)| and \verb|histvars = c('Z1', 'Z2', 'Z3)|, the variables \verb|lag1_Z1|, \verb|lag1_Z2|, \verb|lag1_Z3|, \verb|lag2_Z1|, \verb|lag2_Z2|, \verb|lag2_Z3|, are automatically created and added to the input data set \verb|obs_data| (see Section \ref{histories}) because they are referenced in the model statements in \verb|covmodels|.

Without additional specifications (see Section \ref{additional}), the distributions of time-varying covariates are fit using records for all times $k$ (i.e. under pooled over time models).  Therefore, users will typically want to allow these models to depend on the interval index $k$ as specified by the argument \verb|time_name| (otherwise, the distributions are assumed the same at all $k$). To allow dependence on time, users can include the time variable \verb|time_name|, or any function of this variable that does not need to be created outside of a model statement in \proglang{R}, in the model statement.  For example, in the above sample call, each covariate mean at time $k$ is assumed a quadratic function of $k$ by the inclusion of the terms \verb|t0| and \verb|I(t0^2)|. In Section \ref{discrete_time}, we describe how to allow these distributions to depend on a function of time that cannot be created within a model statement.  

\subsection{Allowing distributions to depend on a categorization of time}\label{discrete_time}
Users may choose to allow the distributions of time-varying covariates to depend on a categorized function of the time index, e.g. indicators for quartiles of time. To implement this, a new variable based on a categorization of \verb|time_name| must first be created and added to the input data set. The naming convention requires that the categorical time variable has the same name as the continuous time variable, with \verb|_f| appended to the end. For example, the call below creates indicators for quartiles of time, which assumes that the input data set has 8 time points and the time index is called \verb|t0|:

\begin{verbatim}
thresholds <- c(1, 3, 5)
basicdata$t0_f <- ifelse(basicdata$t0 <= thresholds[1], 0, 
                         ifelse(basicdata$t0 <= thresholds[2], 1,
                                ifelse(basicdata$t0 <= thresholds[3], 2, 3)))
basicdata$t0_f <- as.factor(basicdata$t0_f)
\end{verbatim}

While the time index, or any function of this index, is conceptually not a time-varying covariate, when pooled over time models are assumed to be a function of categorized time, the time variable with appendix \verb|_f| must be included as a component of the argument \verb|covnames|. When this is included the corresponding component of \verb|covtypes| should be set to \verb|'categorical time'| and the corresponding component of \verb|covmodels| should be set to \verb|NA|.

Sample syntax is as follows:

\begin{verbatim}
gformula_survival(..., time_name = 't0', 
                       covnames = c('L', 'A', 't0_f'),
                       histories = c(lagged),
                       histvars = list(c('A')),
                       covtypes = c('binary', 'binary', 'categorical time'),
                       covparams = list(covmodels = c(L ~ lag1_A + lag1_L + t0_f, 
                                                      A ~ lag1_A + L + lag1_L + t0_f, 
                                                      NA)),
                       )
\end{verbatim}

Note that the modified time variable with extension \verb|_f| need only be created and referenced when fewer categories than levels of the original time variable \verb|time_name| are desired by the user. When users choose to define one category per level of \verb|time_name|, an indicator for each level of the time index may be created within the model statement using \verb|as.factor(time_name)| without additional pre-processing steps as illustrated in Section \ref{examples}.

\subsection{Specifying outcome and competing event models}\label{outcome}
The argument \verb|ymodel| takes as input an \proglang{R} model statement. For survival outcomes this model statement is passed to \verb|glm| in \proglang{R} with  \verb|binomial| family and \verb|logit| link to estimate the hazard at each time $k$ conditional on history (i.e. $p_k(\overline{l}_k,\overline{a}_k)$ from step 1.b of the algorithm in Section \ref{survalgorithm}).  Therefore, this model should generally depend on a function of the time $k$ as specified in \verb|time_name|.  

For continuous end of follow-up outcomes, this model statement is passed to \verb|glm| with \verb|gaussian| family and \verb|identity| link to estimate the mean of the outcome at end of follow-up conditional on history (i.e. $\mu(\overline{l}_K,\overline{a}_K)$ from modified step 2.c of the algorithm in Section \ref{eofalgorithm}. Therefore, this model is not dependent on time $k$.  Similarly, for binary end of follow-up outcomes, this model statement is passed to \verb|glm| with  \verb|binomial| family and \verb|logit| link to estimate the mean of the outcome at end of follow-up conditional on history (which is not dependent on time $k$).

The argument \verb|compevent_model| takes as input an \proglang{R} model statement passed to \verb|glm| in \proglang{R} with \verb|binomial| family and \verb|logit| link to estimate the hazard at each time $k$ conditional on history (i.e. $q_k(\overline{l}_k,\overline{a}_k)$ from step 1.c of the algorithm in Section \ref{survalgorithm}).  The model is generally dependent on time $k$. This argument should only be specified for survival outcomes and when the user chooses not to define the competing event as a censoring event. 

Various syntax examples using the arguments \verb|ymodel| and \verb|compevent_model| are provided in Section \ref{examples}. 

\subsection{Specifying the interventions}\label{strategies}
The arguments \verb|intvars|, \verb|interventions|, and \verb|int_times| are jointly used to define the user-specified treatment interventions/strategies/rules $h^{user}(\overline{a}_k,a^{*}_k,\overline{l}_k)$ to be compared. 

\verb|intvars| is a list of vectors.  The number of vector elements of this list should be the number of user-specified interventions of interest.  Each vector element specifies the time-varying covariates to be intervened upon under that intervention (i.e. the treatments/exposures).  This vector will have a single element if the intervention involves intervening on only a single time-varying covariate (we will call this a \textsl{single intervention}) and will have multiple elements if the intervention involves intervening on multiple time-varying covariates (a \textsl{joint intervention}).

\verb|interventions| is a list whose elements are lists of vectors. The number of list elements of this list should be the number of user-specified interventions of interest.  Each list element in \verb|interventions| specifies one or more intervention rules with each rule defined by a vector of arguments.  The list element will include only a single vector for an element corresonding to a single intervention rule and will include multiple vectors for an element corresponding to a joint intervention rule (each vector element specifying the intervention rule for a given covariate under that intervention). The first element of a vector specifying an intervention rule for a given covariate must be the function that defines the intervention and the following elements specify the input parameter(s) of the intervention function (if applicable). The following is a list of pre-coded intervention rules available with the package with examples (see Section \ref{examples} with additional examples):
\begin{itemize}
    \item \verb|static|: The function \verb|static| specifies a deterministic static intervention.  It requires specification of a vector of length \verb|time_points| with the $k^{th}$ element the value of treatment to be assigned under that intervention rule at time $k$.  The following is sample syntax to compare two static interventions on a binary time-varying covariate \verb|A| that sets this variable to 0 at all times versus 1 at all times:
    \begin{verbatim}
gformula_survival(..., time_points = 10,
                       intvars = list(c('A'), c('A')), 
                       interventions = list(list(c(static, rep(0, 10))), 
                                            list(c(static, rep(1, 10)))))
\end{verbatim}
The following extends this example to compare joint static interventions that sets two time-varying covariates  \verb|A1| and \verb|A2| to 0 at all times versus 1 at all times: 

\begin{verbatim}
gformula_survival(..., time_points = 10,
                       intvars = list(c('A1', 'A2'), c('A1', 'A2')), 
                       interventions = list(list(c(static, rep(0, 10)), 
                                                 c(static, rep(0, 10))), 
                                            list(c(static, rep(1, 10)), 
                                                 c(static, rep(1, 10)))))
\end{verbatim}
    \item \verb|threshold|: The function \verb|threshold| specifies a threshold intervention, an example of an intervention at $k$ that depends on the natural value of treatment $k$ (see Section \ref{estimands}. Under a threshold intervention, at each time $k$, the treatment value is set to the simulated (natural) value of treatment at $k$ if it falls within a user-specified range and, otherwise, the treatment value under intervention is set to the max of that range (if the simulated value is above) or the min of that range (if the simulated value is below) \citep{WHOchap,taubman,youngthreshold}. This rule requires specification of the min and max of the desired range.  If no maximum is desired this should be set to \verb|Inf| and if no minimum is desired this should be set to \verb|-Inf|.   The following is sample syntax to compare two threshold interventions maintaining a time-varying covariate \verb|Z2| at all times at $\geq 2$ versus $\geq 3$:
    \begin{verbatim}
gformula_survival(..., intvars = list(c('Z2'), c('Z2')),
                       interventions = list(list(c(threshold, 2, Inf)), 
                                            list(c(threshold, 3, Inf))))
\end{verbatim}
\item \verb|natural|: The function \verb|natural| specifies the natural course intervention under which the simulated value is assigned at each time $k$.  Risks/means under the natural course are always computed by default and thus this intervention does not need to be specified. 
\end{itemize}

To specify the intervention used as the ``reference'', i.e. the denominator in the risk/mean ratio/difference calculations, set \verb|ref_int| equal to the index of \verb|interventions| in which the desired reference intervention is specified. By default, \verb|ref_int = 0|, or the natural course.

 Optionally, users can specify the time points in which interventions are applied via the \verb|int_times| argument. As with the \verb|intvars| argument, \verb|int_times| is a list whose elements are lists of vectors. Each vector specifies the time points in which the relevant intervention is applied on the corresponding variable in \verb|intvars|. When an intervention is not applied at a time $k$ (not included in \verb|int_times|), the simulated (natural) value is used. By default, all interventions are applied at all time points.  The following extension of the above example compares two interventions on the time-varying covariate \verb|Z2| under which the natural value of treatment is assigned at times $k=0,1$ and, for $k\geq 2$, a threshold intervention is applied that ensures the treatment is assigned values $\geq 2$ versus $\geq 3$:

 \begin{verbatim}
gformula_survival(..., time_points=10, intvars = list(c('Z2'), c('Z2')),
                       interventions = list(list(c(threshold, 2, Inf)), 
                                            list(c(threshold, 3, Inf))),
                       int_times = list(list(2:9), list(2:9)))
\end{verbatim}

\subsection{Bootstrapping and parallelization}\label{bootstrapping}

The \verb|gformula|-type functions can implement a nonparametric bootstrap to estimate standard errors and 95\% confidence intervals. The parameter \verb|nsamples| specifies the number of bootstrap samples.  By default, this parameter is set to 0, indicating no bootstrap samples. See Section \ref{algorithm} for details.

Applications of the g-formula can be computationally expensive, especially when a large number of bootstrap samples are generated. To parallelize bootstrapping and estimation under each intervention, set \verb|parallel = TRUE|. When parallelization is used, users must specify the desired number of CPU cores to parallelize across. In many applications, users may wish to set the number of cores equal to the number of total available cores minus one. 

The following is sample syntax for applying 500 bootstrap replicates of the g-formula in parallel across the total number of available cores minus one: 
\begin{verbatim}
ncores <- parallel::detectCores() - 1
gformula_survival(..., nsamples = 500, parallel = TRUE, ncores = ncores)
\end{verbatim}

\subsection{Output}

The \verb|g-formula|-type functions return S3 class objects, which are lists that contain the following main elements: (1) a \verb|data.table| containing the nonparametric estimates of the natural course risk / mean outcome (see Section \ref{algorithm}) and the parametric g-formula estimates of the risk / mean outcome under each specified intervention at each time point, (2) the coefficients, standard errors, and RMSE values of the models fit in step 1 of the algorithm, and optionally (3) the simulated histories from step 2 of the algorithm under each specified intervention. For the simulated histories to be included in the output, users must set the parameter \verb|sim_data_b = TRUE| in the \verb|g-formula|-type functions.

S3 \verb|print| methods are available for these objects, which have arguments \verb|coefficients|, \verb|stderrs|, \verb|rmses|, and \verb|hazard_ratio| to print the coefficients, standard errors, RMSE values, and hazard ratios (if applicable, see Section \ref{hazard_ratios}), respectively. S3 \verb|plot| methods are also available, which allow the user to plot parametric g-formula versus nonparametric estimates of the risks by each follow-up interval under the natural course intervention for \verb|gformula_survival|-type objects, along with parametric versus nonparametric estimates of covariate means under the natural course for \verb|gformula_survival|, \verb|gformula_continuous_eof|, and \verb|gformula_binary_eof|-type objects.  Also see \cite{pgformyoung}.   Users can specify which covariates to plot, the ordering of plots in the grid, the grid arrangement, and basic plotting options. For \verb|gformula_survival| objects, users can also specify whether to include plots of both risk and survival. See Section \ref{examples} for examples illustrating these S3 methods.

\section{Examples}\label{examples}
In this section, we provide additional examples along with a description of output.  

\subsection*{Example 1: Estimating the effect of static treatment strategies on risk of a failure event}

The example dataset \verb|basicdata_nocomp| consists of 13,170 observations on 2,500 individuals with a maximum of 7 follow-up times. No individuals are censored in this data set. The variables in the dataset are:

\verb|t0|: The time index \\
\verb|id|: A unique identifier for each individual \\
\verb|L1|: A time-varying covariate; binary  \\
\verb|L2|: A time-varying covariate; continuous  \\
\verb|L3|: A baseline covariate; continuous  \\
\verb|A|: The treatment variable; binary \\
\verb|Y|: The outcome/failure event of interest; time-varying indicator of failure

Data for the first subject, who survives the whole follow-up, is presented below:

\begin{verbatim}
   id t0 L1         L2 L3 A Y
1:  1  0  0  1.1470920  5 1 0
2:  1  1  0 -0.9254032  5 1 0
3:  1  2  0 -0.9899824  5 0 0
4:  1  3  1  1.0057421  5 1 0
5:  1  4  1 -1.1956468  5 1 0
6:  1  5  0 -0.9697723  5 1 0
7:  1  6  1 -1.0887002  5 1 0
\end{verbatim}

The subject identifier \verb|id| does not have to be a number, as long as it is unique for each individual.

The following syntax can be used to estimate, using this data set, the risk of failure from the event of interest by $K+1=7$ under a strategy that sets treatment to 0 at all follow-up times for all individuals (``never treat'') and a strategy that sets it to 1 at all times (``always treat'').  The risk under the natural course is automatically estimated. Note that we must use the \verb|nsimul| parameter to set the number of simulated histories ($s$, Section \ref{algorithm}) because in this example the baseline sample size $n<10000$.

\begin{verbatim}
> id <- 'id'
> time_points <- 7
> time_name <- 't0'
> covnames <- c('L1', 'L2', 'A')
> outcome_name <- 'Y'
> covtypes <- c('binary', 'bounded normal', 'binary')
> histories <- c(lagged, lagavg)
> histvars <- list(c('A', 'L1', 'L2'), c('L1', 'L2'))
> covparams <- list(covmodels = c(L1 ~ lag1_A + lag_cumavg1_L1 + lag_cumavg1_L2 +
+                                   L3 + t0,
+                                 L2 ~ lag1_A + L1 + lag_cumavg1_L1 +
+                                   lag_cumavg1_L2 + L3 + t0,
+                                 A ~ lag1_A + L1 + L2 + lag_cumavg1_L1 +
+                                   lag_cumavg1_L2 + L3 + t0))
> ymodel <- Y ~ A + L1 + L2 + L3 + lag1_A + lag1_L1 + lag1_L2 + t0
> intvars <- list('A', 'A')
> interventions <- list(list(c(static, rep(0, time_points))),
+                       list(c(static, rep(1, time_points))))
> int_descript <- c('Never treat', 'Always treat')
> nsimul <- 10000
> 
> gform_basic <- gformula_survival(obs_data = basicdata_nocomp, id = id,
+                                  time_points = time_points,
+                                  time_name = time_name, covnames = covnames,
+                                  outcome_name = outcome_name,
+                                  covtypes = covtypes,
+                                  covparams = covparams, ymodel = ymodel,
+                                  intvars = intvars,
+                                  interventions = interventions,
+                                  int_descript = int_descript,
+                                  histories = histories, histvars = histvars,
+                                  basecovs = c('L3'), nsimul = nsimul,
+                                  seed = 1234)
> gform_basic
PREDICTED RISK UNDER MULTIPLE INTERVENTIONS

Intervention 	 Description
0              Natural course
1              Never treat
2              Always treat

Sample size = 2500, Monte Carlo sample size = 10000
Number of bootstrap samples = 0
Reference intervention = natural course (0)
 

 k Interv. NP risk g-form risk Risk ratio Risk difference
 6       0  0.5056   0.5048278  1.0000000       0.0000000
 6       1      NA   0.7314627  1.4489351       0.2266349
 6       2      NA   0.2339747  0.4634743      -0.2708531
\end{verbatim}

In this example, the parametric g-formula estimates of the risk (\verb|g-form risk|) by end of follow-up under ``never treat'',``always treat'' and the natural course are 0.731, 0.234 and 0.505, respectively.  Because \verb|refint| was not specified, the reference intervention for calculation of risk ratios and risk differences is the natural course.  For example, the risk ratio by $K+1$ comparing ``never treat'' to the natural course is 1.45.   The nonparametric estimate (\verb|NP risk|) of the natural course risk is 0.506 and close to the parametric g-formula estimate of this risk.  This supports (but does not guarantee) the absence of gross model misspecification \citep{pgformyoung}.

In this example, the \verb|plot(gform_basic)| command additionally produces the following default graphs comparing nonparametric and parametric g-formula estimates of the event risk by each follow-up time, and covariate means, under the natural course: 

\begin{center}
	\includegraphics{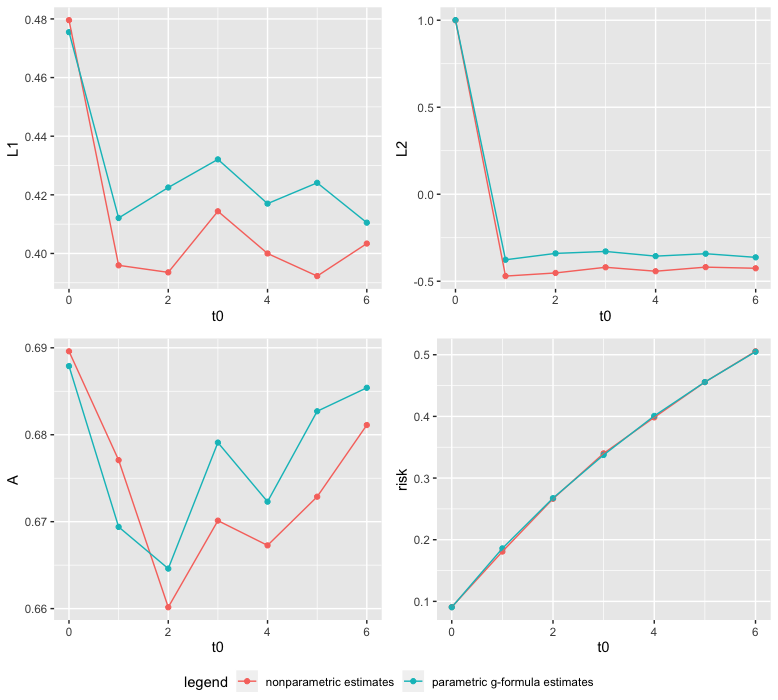}
\end{center}

\subsection*{Example 2: Estimating the effect of treatment strategies on risk of a failure event when competing events exist}

The example dataset \verb|basicdata| consists of 11,332 observations on 2,500 individuals and a maximum of 7 follow-up times. The variables in the dataset are:

\verb|t0|: The time index \\
\verb|id|: A unique identifier for each individual \\
\verb|L1|: A time-varying covariate; binary \\
\verb|L2|: A time-varying covariate; continuous \\
\verb|L3|: A baseline covariate; continuous \\
\verb|A|: The treatment variable; binary \\
\verb|D|: Competing event indicator; time-varying indicator of failure from a competing event \\
\verb|Y|: The outcome; time-varying indicator of failure from the event of interest

Again, we illustrate estimation of the risk of failure from the event of interest by end of follow-up under ``never treat'', ``always treat'' and the natural course.  The difference between this example and the previous one is that here competing events are present such that, once a competing event occurs, an individual cannot subsequently experience the event of interest.  Here we do not treat competing events as censoring events and therefore aim to estimate the total effect of adhering to different strategies, which may capture an effect of the treatment on the competing event \citep{crpaper}.  Below, we illustrate the modified syntax for this setting, including the fitting of pooled over time models as a function of a discrete function of time (see Section \ref{discrete_time}). We also illustrate use of the S3 \verb|print| method to print RMSE values of the fitted models: 

\begin{verbatim}
> id <- 'id'
> time_points <- 7
> time_name <- 't0'
> covnames <- c('L1', 'L2', 'A')
> outcome_name <- 'Y'
> compevent_name <- 'D'
> covtypes <- c('binary', 'bounded normal', 'binary')
> histories <- c(lagged, lagavg)
> histvars <- list(c('A', 'L1', 'L2'), c('L1', 'L2'))
> covparams <- list(covlink = c('logit', 'identity', 'logit'),
+                   covmodels = c(L1 ~ lag1_A + lag_cumavg1_L1 + lag_cumavg1_L2 +
+                                   L3 + as.factor(t0),
+                                 L2 ~ lag1_A + L1 + lag_cumavg1_L1 +
+                                   lag_cumavg1_L2 + L3 + as.factor(t0),
+                                 A ~ lag1_A + L1 + L2 + lag_cumavg1_L1 +
+                                   lag_cumavg1_L2 + L3 + as.factor(t0)))
> ymodel <- Y ~ A + L1 + L2 + lag1_A + lag1_L1 + lag1_L2 + L3 + as.factor(t0)
> compevent_model <- D ~ A + L1 + L2 + lag1_A + lag1_L1 + lag1_L2 + L3 + as.factor(t0)
> intvars <- list('A', 'A')
> interventions <- list(list(c(static, rep(0, time_points))),
+                       list(c(static, rep(1, time_points))))
> int_descript <- c('Never treat', 'Always treat')
> nsimul <- 10000
> 
> gform_basic <- gformula_survival(obs_data = basicdata, id = id,
+                                  time_points = time_points,
+                                  time_name = time_name, covnames = covnames,
+                                  outcome_name = outcome_name,
+                                  compevent_name = compevent_name,
+                                  covtypes = covtypes,
+                                  covparams = covparams, ymodel = ymodel,
+                                  compevent_model = compevent_model,
+                                  intvars = intvars, interventions = interventions,
+                                  int_descript = int_descript,
+                                  histories = histories, histvars = histvars,
+                                  basecovs = c('L3'), nsimul = nsimul,
+                                  seed = 1234)
> print(gform_basic, rmses = TRUE)
PREDICTED RISK UNDER MULTIPLE INTERVENTIONS

Intervention 	 Description
0              Natural course
1              Never treat
2              Always treat

Sample size = 2500, Monte Carlo sample size = 10000
Number of bootstrap samples = 0
Reference intervention = natural course (0)
 

 k Interv.   NP risk g-form risk Risk ratio Risk difference
 6       0 0.4441199   0.4231695   1.000000       0.0000000
 6       1        NA   0.6056894   1.431316       0.1825199
 6       2        NA   0.2118213   0.500559      -0.2113482


 RMSE Values
$L1
[1] 0.4084366

$L2
[1] 0.03726525

$A
[1] 0.4726761

$Y
[1] 2.71106

$D
[1] 3.060926

\end{verbatim}

Note that the primary difference between this example and Example 1 is the specification of the parameters \verb|compevent_model| and \verb|compevent_name|, which together estimate  the conditional hazard of the competing event (see Section \ref{survalgorithm}).  

If, in place of the total effect, the user wished to estimate the treatment effect under elimination of competing events (a controlled direct effect), which is equivalent to treating competing events as censoring events \citep{crpaper}, then the parameters \verb|compevent_model| and \verb|compevent_name| should not be specified.

\subsection*{Example 3: Estimating the effect of threshold interventions on the mean of a binary end of follow-up outcome}

The example dataset \verb|binary_eofdata| consists of 17,500 observations on 2,500 individuals with a maximum of 7 follow-up times. The outcome of interest corresponds to the indicator of an event only in the last interval $K+1$. The variables in the dataset are:

\verb|time|: The time index \\
\verb|id_num|: A unique identifier for each individual \\
\verb|cov1|: A time-varying covariate; binary \\
\verb|cov2|: A time-varying covariate; continuous \\
\verb|cov3|: A baseline covariate; continuous \\
\verb|treat|: The treatment variable; binary \\
\verb|outcome|: The outcome of interest; continuous

In this example, we illustrate use of the function \verb|gformula_binary_eof| to estimate the probability of experiencing the outcome in interval $K+1=7$ under a time-varying threshold intervention that maintains the treatment in all intervals at or above 1, as well as under ``never treat'' and the natural course. We also illustrate the construction of 95\% confidence intervals using, for simplicity, 20 bootstrap samples, as well as parallelization of the bootstrapping and estimation procedures across the total number of available CPU cores minus 1. When increasing the number of bootstrap samples to 500 in this example, the program took approximately 70 minutes to run on a MacBook Air with i5 (1.4 GHz) and 4 GB of RAM when using \proglang{R} version 3.5.2.

\begin{verbatim}
> id <- 'id_num'
> time_name <- 'time'
> covnames <- c('cov1', 'cov2', 'treat')
> outcome_name <- 'outcome'
> histories <- c(lagged, cumavg)
> histvars <- list(c('treat', 'cov1', 'cov2'), c('cov1', 'cov2'))
> covtypes <- c('binary', 'zero-inflated normal', 'normal')
> covparams <- list(covmodels = c(cov1 ~ lag1_treat + lag1_cov1 + lag1_cov2 + cov3 +
+                                   time,
+                                 cov2 ~ lag1_treat + cov1 + lag1_cov1 + lag1_cov2 +
+                                   cov3 + time,
+                                 treat ~ lag1_treat + cumavg_cov1 +
+                                   cumavg_cov2 + cov3 + time))
> ymodel <- outcome ~  treat + cov1 + cov2 + lag1_cov1 + lag1_cov2 + cov3
> intvars <- list('treat', 'treat')
> interventions <- list(list(c(static, rep(0, 7))),
+                       list(c(threshold, 1, Inf)))
> int_descript <- c('Never treat', 'Threshold - lower bound 1')
> nsimul <- 10000
> ncores <- parallel::detectCores() - 1
> 
> gform_bin_eof <- gformula_binary_eof(obs_data = binary_eofdata, id = id,
+                                      time_name = time_name,
+                                      covnames = covnames,
+                                      outcome_name = outcome_name,
+                                      covtypes = covtypes,
+                                      covparams = covparams,
+                                      ymodel = ymodel,
+                                      intvars = intvars,
+                                      interventions = interventions,
+                                      int_descript = int_descript,
+                                      histories = histories, histvars = histvars,
+                                      basecovs = c("cov3"), seed = 1234,
+                                      parallel = TRUE, nsamples = 20,
+                                      nsimul = nsimul, ncores = ncores)
> gform_bin_eof
PREDICTED RISK UNDER MULTIPLE INTERVENTIONS

Intervention 	 Description
0              Natural course
1              Never treat
2              Threshold - lower bound 1

Sample size = 2500, Monte Carlo sample size = 10000
Number of bootstrap samples = 20
Reference intervention = natural course (0)
 

 k Interv. NP mean g-form mean     Mean SE Mean lower 95% CI Mean upper 95% CI
 6       0  0.0988  0.10372823 0.005821318        0.09348999         0.1142943
 6       1      NA  0.09844365 0.009519688        0.08653215         0.1179443
 6       2      NA  0.09353075 0.015585168        0.07276734         0.1268941
 Mean ratio      MR SE MR lower 95% CI MR upper 95% CI Mean difference
  1.0000000 0.00000000       1.0000000        1.000000     0.000000000
  0.9490536 0.07795572       0.8054892        1.091798    -0.005284583
  0.9016904 0.14185846       0.6590487        1.175723    -0.010197477
       MD SE MD lower 95% CI MD upper 95% CI
 0.000000000      0.00000000     0.000000000
 0.008379784     -0.02150494     0.009788027
 0.015223406     -0.03769290     0.018737856
\end{verbatim}

The output reports parametric g-formula estimates of the mean outcome (\verb|g-form mean|) under ``never treat'' (0.098), the threshold intervention (0.094) and the natural course (0.104).  It also reports nonparametric estimates of the outcome mean (\verb|NP mean|) under the natural course (0.099), along with 95\% confidence intervals for these means and mean differences and ratios comparing each intervention to the natural course (because \verb|refint|) was not specified).

\subsection*{Example 4: Estimating the effect of treatment strategies on the mean of a continuous end of follow-up outcome}

The example dataset \verb|continuous_eofdata| again consists of 7,500 observations on 2,500 individuals with a maximum of 7 follow-up times, where the outcome corresponds to a characteristic only in the last interval (e.g. systolic blood pressure in interval 7). The variables in the dataset are:

\verb|t0|: The time index \\
\verb|id|: A unique identifier for each individual \\
\verb|L1|: A time-varying covariate; categorical \\
\verb|L2|: A time-varying covariate; continuous \\
\verb|L3|: A baseline covariate; continuous \\
\verb|A|: The treatment variable; binary \\
\verb|Y|: The outcome of interest; continuous

In this example, we illustrate how to estimate the mean outcome at $K+1=7$ under ``never treat'' versus ``always treat''. We also illustrate in this example how to include a restricted cubic spline function of a variable  in a model statement using the \verb|rcspline.eval| function from the \pkg{Hmisc} package:

\begin{verbatim}
> library("Hmisc")
> id <- 'id'
> time_name <- 't0'
> covnames <- c('L1', 'L2', 'A')
> outcome_name <- 'Y'
> covtypes <- c('categorical', 'normal', 'binary')
> histories <- c(lagged)
> histvars <- list(c('A', 'L1', 'L2'))
> covparams <- list(covmodels = c(L1 ~ lag1_A + lag1_L1 + L3 + t0 +
+                                   rcspline.eval(lag1_L2, knots = c(-1, 0, 1)),
+                                 L2 ~ lag1_A + L1 + lag1_L1 + lag1_L2 + L3 + t0,
+                                 A ~ lag1_A + L1 + L2 + lag1_L1 + lag1_L2 + L3 + t0))
> ymodel <- Y ~ A + L1 + L2 + lag1_A + lag1_L1 + lag1_L2 + L3
> intvars <- list('A', 'A')
> interventions <- list(list(c(static, rep(0, 7))),
+                       list(c(static, rep(1, 7))))
> int_descript <- c('Never treat', 'Always treat')
> nsimul <- 10000
> 
> gform_cont_eof <- gformula_continuous_eof(obs_data = continuous_eofdata,
+                                           id = id,
+                                           time_name = time_name,
+                                           covnames = covnames,
+                                           outcome_name = outcome_name,
+                                           covtypes = covtypes,
+                                           covparams = covparams, ymodel = ymodel,
+                                           intvars = intvars,
+                                           interventions = interventions,
+                                           int_descript = int_descript,
+                                           histories = histories, histvars = histvars,
+                                           basecovs = c("L3"),
+                                           nsimul = nsimul, seed = 1234)
> gform_cont_eof
PREDICTED RISK UNDER MULTIPLE INTERVENTIONS

Intervention 	 Description
0              Natural course
1              Never treat
2              Always treat

Sample size = 2500, Monte Carlo sample size = 10000
Number of bootstrap samples = 0
Reference intervention = natural course (0)
 

 k Interv.   NP mean g-form mean Mean ratio Mean difference
 6       0 -4.414543   -4.348234   1.000000       0.0000000
 6       1        NA   -3.107835   0.714735       1.2403991
 6       2        NA   -4.603006   1.058592      -0.2547717
\end{verbatim}

\section{Additional features}\label{additional}

\subsection{Incorporating deterministic knowledge of the data structure}\label{deterministic}

Sometimes we have information about relationships between time-varying covariates that can be used in place of arbitrary parametric model assumptions or other methods of smoothing to avoid extrapolation where unnecessary. For example, consider the case where we have two time-varying covariates corresponding, respectively, to indicators of whether an individual has started menopause by a given interval $k$ (\verb|menopause|) and whether she is pregnant in interval $k$ (\verb|pregnancy|). In this case, we know that given \verb|menopause == 1|, the probability that  \verb|pregnancy == 0| is 1 regardless of other history and can incorporate this knowledge into the algorithm.  

Generally, let $Z_{j,k}$ (e.g. pregnancy in interval $k$)  be a component of the covariate vector $Z_k$ for which we have deterministic knowledge of its distribution given its `history'' $(Z_{j-1, k}, \dots, Z_{1, k}, \overbar{L}_{k-1}, \overbar{A}_{k-1})$ (e.g. menopause status by interval $k$ defined as a $Z_{h,k}$ where $h<j$).  Note this implies a preferred permutation of the components of $Z_k$, rather than an arbitrary permutation (see Section \ref{algorithm}); specifically, in this example, menopause should precede pregnancy in the chosen permutation.  We can incorporate this knowledge as follows into the estimation algorithm described in Section \ref{algorithm}: 
\begin{itemize}
\item In step 1.a of the algorithm restrict the chosen method of estimating the mean of $Z_{j,k}$ (e.g. pregnancy status) given ``history'' to only records where deterministic knowledge is absent.  In our example, this would be the case for records with \verb|menopause == 0|.
\item In step 2.a of the algorithm, set $Z_{j,k}$ deterministically to its known value for histories under which this value is known.  Otherwise, draw $Z_{j,k}$ according to the model-based estimate (or otherwise estimated) conditional distribution of $Z_{j,k}$.  In our example, if the value of \verb|menopause| in step 2.a. at time $k$ is 1 then \verb|pregnancy| is set to 0.  Otherwise, the value of \verb|pregnancy| at time $k$ is drawn from the estimated distribution in step 1.a.
\end{itemize}

The \pkg{gfoRmula} package allows for this type of restricted modeling and modified simulation step using the \verb|restrictions| parameter. To implement the scenario described above, the user will define the following parameters within the the function call:

\begin{verbatim}
gformula_survival(..., restrictions = list(c('pregnancy', 'menopause == 0', 
                                             simple_restriction, 0)))
\end{verbatim}

Note that \verb|restrictions| is a list and can contain multiple vectors; that is, the user can impose multiple modeling restrictions.

For each vector in \verb|restrictions|:
\begin{itemize}
	\item The first entry is the covariate $Z_{j,k}$ (e.g. \verb|pregnancy|) for which we have knowledge of its distribution given a particular ``history.''
	\item The second entry is the condition that must be \emph{true} for the conditional mean of the covariate in the first entry to be modeled; equivalently, the condition under which we do \textsl{not} have knowledge of the distribution of  $Z_{j,k}$ (e.g. \verb|menopause == 0|)
	\item The third entry is a function that determines the value of the covariate in the first entry when the condition in the second entry is \textsl{not} true. In the example above, the \verb|simple_restriction| function simply assigns \verb|pregnancy| a static value when  \\ \verb|menopause != 0|.
	\item The fourth entry is a value used by the function in the third entry. In the above example, this is the value to be assigned to \verb|pregnancy| when \verb|menopause != 0|.
	\item It is also possible to include additional entries; all of these should be values used by the function in the third entry.
\end{itemize}

In addition to \verb|simple_restriction|, the package includes an additional carry-forward restriction type named \verb|carry_forward|. Rather than assign a known value, if the condition in the second entry of \verb|restrictions| is not true, \verb|carry_forward| assigns the value of the covariate at the previous time point (that is, it "carries forward" the value from the previous time point if modeling does not occur at the current time point). For example, in some cases certain covariates will only be measured in certain intervals.  In this case, if (i) $Z_{j,k}$ is defined as the last measured value of a covariate relative to time $k$, and (ii) the input data set is constructed so that at times $k$ where the covariate is not measured then $Z_{j,k}=Z_{j,k-1}$ then we have deterministic knowledge of the distribution of $Z_{j,k}$ given $Z_{j,k-1}$ at those times.  For example, suppose a covariate \verb|smoking| corresponds to the last measured value of smoking status and is only measured in intervals $k=0$, 3, and 6.  To incorporate the fact that the last measured value of smoking status at $k\neq 0,3,6$ must be equal to the value at $k-1$ at those times, we incorporate the following syntax:

\begin{verbatim}
gformula_survival(..., time_name = 't0', 
                       restrictions = list(c('smoking', 't0%in%c(0, 3, 6)',
                                             carry_forward)))
\end{verbatim}
Note that this approach, of course, requires that exchangeability assumptions for the causal effect of interest hold given only last measured values of certain covariates (see Section \ref{assumptions}).  

Simple restrictions may also be placed on the outcome hazard/mean (and the competing event hazard, allowed for survival outcomes only). \verb|yrestrictions|, the parameter for specifying restrictions on the outcome mean/hazard, restricts the estimation for the outcome probability in step 1 to records meeting a specified condition. Then in step 2, if the condition is met based on the previously simulated data, a model-based estimate of the conditional hazard/mean is used. Otherwise, if the condition is not met, the mean/hazard is set to a specified value. (\verb|compevent_restrictions| operates analogously for the competing event.) Restrictions on the outcome or competing event may be used in conjunction with covariate restrictions.

As an example, suppose the user is interested in estimating the risk of ventilator-associated pneumonia by 30-day follow-up. The criteria for this outcome includes being on the ventilator for at least 3 days. Suppose the user is interested in effects on risk of ventilator-associated pneuomonia in a population of ventilated patients of treatment interventions beginning on the first day of ventilation, with $k$ indexing days.  In this case, the event of interest $Y_{k+1}$ cannot take the value 1 until $k=3$ such that the hazard in the first two intervals of follow-up is 0 by definition. To incorporate this knowledge, the function call would include:

\begin{verbatim}
gformula_survival(..., time_name = 'time', yrestrictions = list(c('time>2', 0)))
\end{verbatim}

As with \verb|restrictions|, the \verb|yrestrictions| and \verb|compevent_restrictions| parameters each accept a list of vectors, meaning that it is possible to impose multiple restrictions on the simulation of each variable by including multiple vectors within each list.

\subsection{Attaching a visit process to a covariate}
In clinical cohorts, the data are not usually recorded at regular intervals but rather are recorded every time the patient visits the clinic. Therefore, the times at which the time-varying covariates are measured will vary by subject. In this setting, it is typical to construct the data such that at a time when there is no visit/measurement, the last measured value of a covariate is carried forward (such that $Z_{j,k}$ actually corresponds to the last measured value of a covariate). Furthermore, the user may choose to censor a subject if he or she is not seen after a specified consecutive number of time intervals.  Following \cite{obsplans}, under such measurement processes, the visit process history through $k$ generally should itself be considered part of the confounder history ($\overline{L}_k$).  

Following \cite{pgformyoung}, in these settings where (i) the visit/measurement process is itself defined as a time-varying confounder (i.e., is conceptually a component of $L_k$); (ii) the input data set is constructed such that last measured values are carried forward at times of no new measurements; and (iii) the user constructs the input data set such that a subject is censored after a certain number of consecutive times with no visit/measurement, there are certain deterministic relationships in the data that can be incorporated into both step 1 and step 2 of the parametric g-formula estimation algorithm to potentially reduce model misspecification and increase precision.  First, by (ii), we know that if there is no visit/measurement in interval $k$ then $Z_{j,k}=Z_{j,k-1}$ (i.e., $Z_{j,k}$ is always equal to its lagged value). We also know that, by (iii), for $s$ the max number of consecutive missed visits since the last visit allowed before a subject is censored, the indicator of a visit at time $k$ (itself a component of $L_k$), by (ii) must be 1 if the number of consecutive missed visits since the last visit at $k-1$ is $s$.  The determinisms described in (ii) and (iii) above can in principle be incorporated into the algorithm via the parameters described in Section \ref{deterministic} and \ref{customhistory}. 

Alternatively, the deterministic knowledge in (ii) and (iii) can be incorporated via the parameter \verb|visitprocess|, a list parameter taking vector entries. The first element of each vector gives the name of a time-varying indicator in the input data set of whether a covariate was measured in interval $k$. The second entry gives the name of that covariate. The conditional mean of the covariate given the ``history'' in step 1 will be estimated only for records where there is a current visit. If the visit indicator equals 1, then in step 2, the value of the dependent covariate will be generated from a distribution based on this estimate; otherwise, the last value is carried forward. The third parameter is the max number $s$ of missed measurements of this covariate allowed since the last measurement before a subject is censored. The probability of a visit given the history in step 1 of the algorithm is estimated only using records where the sum of consecutive missed visits through $k-1$ is less than $s$. In step 2, if the sum of consecutive missed visits through $k-1$ is less than $s$, then the visit indicator is simulated from a distribution based on this estimate; otherwise, the visit indicator is set to 1 so as to eliminate subjects with more than $s$ consecutive missed visits.

In the following example, a time-varying covariate in the input data set, \verb|visit|, indicates whether CD4 cell count and viral load lab measurements were taken in interval $k$, and \verb|cd4| and \verb|rna| are the respective covariates for CD4 cell count and viral load. The input data set is constructed such that a subject is censored after 3 consecutive missed lab measurements (i.e. \verb|visit == 0| 3 intervals in a row). In this case the call will include: 
\begin{verbatim}
gformula_survival(..., covnames = c('visit', 'cd4', 'rna'),
                       covtypes = c('binary', 'normal', 'normal'),
                       visitprocess = list(c('visit', 'cd4', 3),
                                           c('visit', 'rna', 3)))
\end{verbatim}
Note that \verb|visit| itself must be included in \verb|covnames| and assigned a value in \verb|covtypes|. 

The above example assumes that CD4 cell count and viral load are either both measured or both not measured at each visit. The following call alternatively allows a data set with separate visit process indicators for CD4 cell count and viral load and separate maximum missed visits determining censoring.
\begin{verbatim}
gformula_survival(..., covnames = c('visit_cd4', 'cd4', 'visit_rna', 'rna'),
                       covtypes = c('binary', 'normal', 'binary', 'normal'),
                       visitprocess = list(c('visit_cd4', 'cd4', 4),
                                           c('visit_rna', 'rna', 5)))
\end{verbatim}
Note in the current version of the package, a covariate attached to a visit process must be measured (i.e. the indicator of measurement must be 1) at baseline for all individuals in the data set.

\subsection[Specifying a custom covariate distribution]{Specifying a custom \normalfont\texttt{covtype}}\label{customcovtype}

In addition to the existing \verb|covtypes|, users are also able to choose their own covariate distributions and methods of estimating their parameters by setting the \verb|covtype| equal to \verb|'other'| and supplying their own fit and prediction functions through the parameters \verb|covfits_custom| and \verb|covpredict_custom|, respectively.

A user-written fit function must take the parameters \verb|covparams| (see Section \ref{covdist}), \verb|covname| (the name of the covariate), \verb|obs_data| (the name of the input data set), and \verb|j| (the index of the covariate). The template for a custom fit function is:
\begin{verbatim}
fit_custom <- function(covparams, covname, obs_data, j){
  covmodels <- covparams$covmodels
  otherparam1 <- covparams$otherparam1
  otherparam2 <- covparams$otherparam2
  fit <- fitfunc(formula = as.formula(paste(covmodels[j])), a = otherparam1[j], 
                 b = otherparam2[j])
  return (fit)
}
\end{verbatim}

A user-written prediction function must take the parameters \verb|obs_data| (the observed data), \verb|newdf| (the simulated dataset at time \verb|t|), \verb|fitcov| (the model object for the covariate), \verb|time_name| (the name of the time variable), \verb|t| (the current time index), \verb|condition| (any condition restricting the modeling of the covariate), \verb|covname| (the name of the covariate), and any number of additional parameters, which are specified in the original \verb|gformula| function call. An example format is given below. Suppose the inputs to \verb|...| are \verb|a = 1|, \verb|b = 0.5|:

\begin{verbatim}
predict_custom <- function(obs_data, newdf, fitcov, time_name, t, condition, 
                           covname, ...){
  n <- dim(newdf)[1]
  theta <- predict(fitcov, type = 'response', newdata = newdf)
  extra_args <- list(...)
  a <- extra_args$a
  b <- extra_args$b
  prediction <- predictfunc(n = n, theta = theta, a = a, b = b)
  return (prediction)
}
\end{verbatim} 

Like all sub-parameters corresponding to the covariates, \verb|covfits_custom| and \verb|covpredict_custom| are vectors of the same length and in the same order as \verb|covnames|. At the index where the custom \verb|covtype| is desired, \verb|covfits_custom| and \verb|covpredict_custom| will contain the names of the user-supplied fit and prediction functions. Multiple custom covtypes can be specified as well.

For example, suppose the user wishes to model a covariate using random forests. Then their respective fit and predict function might look like:

\begin{verbatim}
fit_rf <- function(covparams, covname, obs_data, j){
  covmodels <- covparams$covmodels
  importance <- covparams$importance
  maxnodes <- covparams$maxnodes
  ntree <- covparams$ntree
  fit <- randomForest::randomForest(formula = as.formula(paste(covmodels[j])),
                                    data = obs_data, importance = importance[j],
                                    maxnodes = maxnodes[j], ntree = ntree[j])
  return (fit)
}

predict_rf <- function(obs_data, newdf, fitcov, time_name, t, condition, 
                       covname, ...){
  extra_args <- list(...)
  proximity <- extra_args$proximity
  prediction <- predict(object = fitcov, newdata = newdf,
                        proximity = proximity)
  return (prediction)
}

\end{verbatim}

 Here, \verb|fit_rf| requires the sub-parameters \verb|importance|, \verb|maxnodes|, and \verb|ntree|. Likewise, \verb|predict_rf| requires the parameter \verb|proximity|. Therefore, the call to the gformula should follow the format:

\begin{verbatim}
gformula_survival(..., covnames = c('L1', 'L2', 'A'),
                       covtypes = c('binary', 'custom', 'zero-inflated normal'),
                       covparams = list(covmodels = c(L1 ~ lag1_L1 + lag1_L2 + 
                                                        lag1_A,
                                                      L2 ~ L1 + lag1_L1 +lag1_L2 + 
                                                        lag1_A, 
                                                      A ~ L1 + lag1_L1 + L2 + 
                                                        lag1_L2 + lag1_A),
                                        covlink = c('logit', NA, 'identity'),
                                        importance = c(NA, TRUE, NA),
                                        maxnodes = c(NA, 4, NA),
                                        ntree = c(NA, 30, NA)),
                       covfits_custom = c(NA, fit_rf, NA),
                       covpredict_custom = c(NA, predict_rf, NA),
                       proximity = FALSE)
\end{verbatim}

To use custom covariate fit and predict functions in conjunction with the \verb|parallel| option, users should take care to load any packages used by those custom functions into the global environment using \verb|library(<packagename>)| prior to calling the \verb|gformula| function, as the packages will not be successfully exported to the clusters otherwise.

\subsection{Specifying a custom history}\label{customhistory}

If users wish to generate histories other than the three provided by the package, they may write those functions themselves. These functions take parameters \verb|pool| (a \verb|data.table| containing the simulated data up until the current time point \verb|t|), \verb|histvars| (the name of the time-varying covariates for which functions of histories will be created), \verb|time_name| (the name of the time variable), \verb|t| (the current time index), and \verb|id_name| (the name of the ID variable). A history function generates in \verb|pool| the value of the desired function of history of each covariate listed in \verb|histvars| at time \verb|t| based on the covariates values at times prior to (and, possibly, at time) \verb|t|. The function does not return a new dataframe; rather, it modifies \verb|pool| in place to reduce memory use.  No object should be returned by a custom history function. 

An example function that generates the average of the three most recent values (i.e., the current value and the last two lagged values for $t>1$, the current value and the last lag for $t=1$ and the current value for $t=0$) is given below:

\begin{verbatim}
ave_last3 <- function(pool, histvars, time_name, t, id_name){
  i <- min(c(t, 2))
  
  # Get indicators for individuals in the (observed or simulated) data set at time t
  current_ids <- unique(pool[get(time_name) == t][[id_name]])
  
  # At time t, for each element (histvar) in histvars, create / update 
  # the column called ave_last3_<histvar> which contains the average of 
  # the three most recent values of histvar
  lapply(histvars, FUN = function(histvar){
    pool[get(time_name) == t,
         (paste("ave_last3_", histvar, sep = "")) :=
           as.double(tapply(pool[get(id_name) %in% current_ids &
                                   get(time_name) <= t &
                                   get(time_name) >= t-i][[histvar]],
                            pool[get(id_name) %in% current_ids &
                                   get(time_name) <= t &
                                   get(time_name) >= t-i][[id_name]],
                            FUN = sum) / (i + 1))]
  })
}
\end{verbatim}

In general, users will have to create an indicator for individuals in the data set at time $t$, as illustrated in the above example. This is because the custom history function must work not only on the simulated data, which contains $K+1$ records for every value of \verb|id|, but also on observed data, which may include fewer records per \verb|id|. In this case, custom history functions will typically use such an indicator to obtain the past covariate values for the individuals in the data set at time $t$. 

To use this newly written function in a \verb|gformula| call, the user may write, for example:
\begin{verbatim}
gformula_survival(..., covnames = c('L1', 'L2', 'A'),
                  covparams = list(covmodels = c(L1 ~ lag1_L1 + lag1_L2 +
                                                   lag1_A + t0,
                                                 L2 ~ L1 + lag1_L1 + lag1_L2 +
                                                   lag1_A + t0,
                                                 A ~ ave_last3_L1 +
                                                   ave_last3_L2 + lag1_A + t0)),
                  histories = c(lagged, ave_last3),
                  histvars = list(c('L1', 'L2', 'A'), c('L1', 'L2')))
\end{verbatim}

\subsection{Specifying a custom intervention}\label{customint}

Beyond the two interventions provided, \verb|static| and \verb|threshold|, users may also write their own intervention functions. These must accept the parameters \verb|newdf| (a \verb|data.table| containing the simulated dataset at time \verb|t|), \verb|pool| (a \verb|data.table| containing the simulated dataset at times prior to \verb|t|, i.e. from \verb|0, ..., t - 1|), \verb|intvar| (the name of the time-varying covariate to be intervened on), \verb|intvals| (a list of one or more values needed internally by the intervention function), \verb|time_name| (the name of the time variable), and \verb|t| (the current time index). 

Several technical considerations should be made when writing custom interventions functions. 
\begin{itemize}
	\item As with custom history functions, no objects should be returned by custom intervention functions. Instead, custom intervention functions should modify \verb|newdf| by reference (i.e., using the \verb|:=| operator from \verb|data.table|). The following examples illustrate how to add and update columns by reference for a custom intervention. 
	\item \verb|newdf| is initiated to the simulated data set at time $t-1$. Prior to calling the custom intervention function, the values of the time-varying covariates are updated to their simulated values at time $t$. Consequently, any auxiliary columns added to \verb|newdf| (e.g., columns \verb|cond_met_ever| and \verb|cond_tracker| in the example \verb|dyn_int| below) are initially set to their values at time $t-1$. Users should therefore only modify auxiliary columns if their values need to be updated from time $t-1$. Moreover, the value of the treatment covariate at time $t$ is initially set to its simulated value under the natural course intervention. Therefore, users should not update the value of the treatment covariate in cases where the natural course is desired. 
\end{itemize}

We now give two examples of user-defined intervention functions that implement dynamic interventions; interventions that depend on time-updated values of covariates:
\subsubsection{Example 1: a dynamic intervention based on the current value of a time-varying covariate}

The following is an example of a function that defines an intervention under which treatment is given only for individuals where \verb|L2| is above a certain threshold (provided in \verb|intvals|):

\begin{verbatim}
example_intervention <- function(newdf, pool, intvar, intvals, time_name, t){
  newdf[(intvar) := 0]
  newdf[L2 < intvals[[1]], (intvar) := 1]
}
\end{verbatim}

Supposing the user wants to set the threshold to 0.75, the function call would then be:

\begin{verbatim}
gformula_survival(..., intvars = list('A'), 
                       interventions = list(list(c(example_intervention, 0.75))))
\end{verbatim}

\subsubsection{Example 2: a (random) dynamic intervention based on a function of the history of a time-varying covariate}

The following is an example of a function that defines a more complex dynamic intervention ``start combined antiretroviral therapy (cART) within $m$ months if CD4 cell count first drops below $x$'' for some grace period of $m>0$ months (assuming intervals $k$ are one month long) and for some cut off $x$ \citep{pgformyoung,cainijb}. During the grace period, treatment is assigned according to the observed distribution of treatment given an individual's past measured treatment and confounder history.  In the data set, the column \verb|lncd4| corresponds to the natural logarithm of the CD4 cell count and the column \verb|art| corresponds to an indicator of cART  initiation (\verb|art=1| indicates treatment initiation).

\begin{verbatim}
dyn_int <- function(newdf, pool, intvar, intvals, time_name, t){
  threshold <- intvals[[1]]
  m <- intvals[[2]]

  # Determine whether threshold has ever been met and track when it first occurs
  if (t == 0){
    newdf[lncd4 < threshold, `:=` (cond_met_ever = 1, cond_tracker = t)]
    newdf[lncd4 >= threshold, cond_met_ever := 0]
  } else {
     newdf[cond_met_ever == 0 & lncd4 < threshold, `:=`
     (cond_met_ever = 1, cond_tracker = t)]
  }
  # If treatment has been initiated by time t-1, then it has been initiated by time t
  if (t > 0){
    newdf[pool[get(time_name) == (t - 1), get(intvar) == 1], (intvar) := 1]
  }

  # If threshold has never been met, set treatment to 0.
  newdf[cond_met_ever == 0, (intvar) := 0]

  # If threshold was met at time t-m, set treatment at t to 1
  if (t >= m){
    newdf[cond_tracker <= (t - m), (intvar) := 1]
  }
}
\end{verbatim}

To set a threshold of 350 for CD4 cell count and a grace period of $m=6$ months, the function call would be:

\begin{verbatim}
gformula_survival(..., intvars = list('art'), 
                       interventions = list(list(c(dyn_int, log(350), 6))))
\end{verbatim}

To verify that the custom intervention function is working properly, users should inspect the simulated data set under the intervention. For instance, below are simulated data for the first ten time points for two ``individuals'' (or, more precisely, simulated histories) under this random dynamic treatment regime:
\begin{verbatim}
    month art    lncd4 cond_met_ever cond_tracker
 1:     0   0 6.142037             0           NA
 2:     1   0 5.973712             0           NA
 3:     2   0 5.885895             0           NA
 4:     3   0 5.694269             1            3
 5:     4   1 5.732024             1            3
 6:     5   1 5.620208             1            3
 7:     6   1 5.485906             1            3
 8:     7   1 5.666780             1            3
 9:     8   1 5.479236             1            3
10:     9   1 5.514820             1            3
\end{verbatim}

The simulated data above is consistent with the desired intervention rule because the condition to start treatment is met at time (\verb|month|) 3 and treatment (\verb|art|) is started at time 4 which is within the grace period (6 months) of meeting the condition to start.

\begin{verbatim}
    month art    lncd4 cond_met_ever cond_tracker
 1:     0   0 5.669881             1            0
 2:     1   0 5.874165             1            0
 3:     2   0 5.603337             1            0
 4:     3   0 5.699042             1            0
 5:     4   0 5.850501             1            0
 6:     5   0 5.775751             1            0
 7:     6   1 5.719197             1            0
 8:     7   1 5.765443             1            0
 9:     8   1 5.695890             1            0
10:     9   1 5.859916             1            0
\end{verbatim}
The simulated data above is consistent with the desired intervention rule because the condition to start treatment is met at time 0 and treatment is started at time 6, the end of the grace period.

\subsection{Hazard ratios} \label{hazard_ratios}

For survival outcomes only (using the function \verb|gformula_survival|), users have the options of calculating the hazard ratio comparing two interventions using the argument \verb|intcomp|. If, for example, the user wishes to calculate the hazard ratio between the natural course and the second intervention specified in the parameter \verb|interventions|, then \verb|intcomp| should be set as \verb|intcomp = list(0, 2)|. When bootstrapping is used (i.e., by setting \verb|nsamples > 0|), a 95\% confidence interval for the hazard ratio is then also computed. Note that the hazard ratio calculation is available only for survival analysis.  When competing events are present in the data and the parameters \verb|compevent_model| and \verb|compevent_name| are specified, the \textsl{subdistribution} hazard ratio \citep{finegray} is computed.  Otherwise, if these are not specified, a quantity that, depending on underlying assumptions, can be interpreted as either a \textsl{marginal} or \textsl{cause-specific} hazard ratio is computed \citep{kalbprentice,crpaper}.  We include these hazard ratios as options for users but generally discourage reporting these as causal effect measures because even counterfactual contrasts in any of these versions of the hazard ratio generally do not have a causal interpretation \citep{Hernanfail,hazards,crpaper}.  

\subsection[Summary of key arguments to the gformula functions]{Summary of key arguments to the \texttt{gformula} functions}

\begin{itemize}
	\item \verb|id|: Character string specifying the name of the ID variable in \verb|obs_data|. 
	\item \verb|time_points|: Specified end of follow-up.
	\item \verb|obs_data|: Data table containing the observed data.
	\item \verb|seed|: Starting seed for simulations and bootstrapping.
	\item \verb|nsimul|: Number of simulated covariate histories under intervention. By default, this argument is set equal to the number of subjects in \verb|obs_data| at baseline (the number of records with \verb|time_name=0|.
	\item \verb|time_name|: Character string specifying the name of the variable indexing follow-up time in \verb|obs_data| (must begin at 0 and increment by 1). 
	\item \verb|outcome_name|: Character string specifying the name of the outcome variable in \verb|obs_data|.
	\item \verb|compevent_name|: Character string specifying the name of the competing event variable in \verb|obs_data|. Can be used only in \verb|gformula_survival|.
	\item \verb|intvars|: List, whose elements are vectors of character strings. The $j$th vector in \verb|intvars| specifies the name(s) of the variable(s) to be intervened on under the $j$th intervention in \verb|interventions|.
	\item \verb|interventions|: List, whose elements are lists of vectors. Each list in \verb|interventions| specifies a unique intervention on the corresponding variable(s) in \verb|intvars|. Each vector contains a function implementing a particular intervention on a single variable, with the first element of the vector always the name of the function and possible subsequent elements requiring integer values or vectors needed by the function.
	\item \verb|int_descript|: Vector of character strings, each describing an intervention. It must be in same order as the entries in \verb|interventions|.
	\item \verb|ref_int|: Integer denoting the intervention to be used as the reference for calculating the risk ratio (in \verb|gformula_survival|) or the end-of-follow-up mean ratio (in \verb|gformula_binary_eof| or \verb|gformula_continuous_eof|). 0 denotes the natural course, while subsequent integers denote user-specified interventions in the order that they are named in \verb|interventions|. The default is 0.
	\item \verb|covnames|: Vector of character strings specifying the names of the time-varying covariates in \verb|obs_data|.
	\item \verb|covtypes|: Vector of character strings specifying the "type" of each time-varying covariate included in \verb|covnames|. The possible "types" are: \verb|"binary"|, \verb|"normal"|, \verb|"categorical"|, \verb|"bounded normal"|, \verb|"zero-inflated normal"|, \verb|"truncated normal"|,  \verb|"absorbing"|, and \verb|``categorical time''|.
	\item \verb|covparams|: List of vectors, where each vector contains information for one parameter used in the modeling of the time-varying covariates (e.g., model statement, family, link function, etc.). Each vector must be the same length as \verb|covnames| and in the same order. If a parameter is not required for a certain covariate, it should be set to \verb|NA| at that index.
	\item \verb|covfits_custom|: Vector containing custom fit functions for time-varying covariates other than the pre-defined covariate types. It should be in the same order as 
	\verb|covnames|. If a custom fit function is not specified for a particular covariate (e.g., if the first covariate is of type \verb|"binary"| but the second is of type \verb|"custom"|), then that index should be set to \verb|NA|. The default is \verb|NA|.
	\item \verb|covpredict_custom|: Vector containing custom prediction functions for time-varying covariates other than the pre-defined covariate types. It should be in the same order as \verb|covnames|. If a custom prediction function is not required for a particular covariate, then that index should be set to \verb|NA|. The default is \verb|NA|.
	\item \verb|histvars|: Vector of character strings specifying the names of the variables for which history functions are to be applied. The default is \verb|NA|.
	\item \verb|histories|: Vector of history functions to apply to the variables specified in \verb|histvars|. The default is \verb|NA|.
	\item \verb|ymodel|: Model statement for the outcome variable.
	\item \verb|yrestrictions|: List of vectors. Each vector contains as its first entry a condition and its second entry an integer. When the condition is \verb|TRUE|, the outcome hazard/mean is estimated according to the fitted model; when the condition is \verb|FALSE|, the outcome hazard/mean takes on the value in the second entry. The default is \verb|NA|.
	\item \verb|compevent_restrictions|: List of vectors. Each vector containins as its first entry a condition and its second entry an integer. When the condition is \verb|TRUE|, the competing event hazard is estimated according to the fitted model; when the condition is \verb|FALSE|, the competing event hazard takes on the value in the second entry. The default is \verb|NA|. Can be used only in \verb|gformula_survival|.
	\item \verb|restrictions|: List of vectors. Each vector contains as its first entry a covariate for which \textsl{a priori} knowledge of its distribution is available; its second entry a condition under which no knowledge of its distribution is available and that must be \verb|TRUE| for the distribution of that covariate given that condition to be estimated via a parametric model or other fitting procedure; its third entry a function for estimating the distribution of that covariate given the condition in the second entry is false such that \textsl{a priori} knowledge of the covariate distribution is available; and its fourth entry a value used by the function in the third entry. The default is \verb|NA|.
	\item \verb|visitprocess|: List of vectors. Each vector contains as its first entry the covariate name of a visit process; its second entry the name of a covariate whose modeling depends on the visit process; and its third entry the maximum number of consecutive visits that can be missed before an individual is censored. The default is \verb|NA|.
	\item \verb|compevent_model|: Model statement for the competing event variable. The default is \verb|NA|. Can be used only in \verb|gformula_survival|.
	\item \verb|intcomp|: List of two numbers indicating a pair of interventions to be compared by a hazard ratio. The default is \verb|NA|, resulting in no hazard ratio calculation. Can be used only in \verb|gformula_survival|.
	\item \verb|sim_data_b|: Logical scalar indicating whether to return the simulated data set. If bootstrap samples are used (i.e., nsamples is set to a value greater than 0), this argument must be set to \verb|FALSE|. The default is \verb|FALSE|.
	\item \verb|nsamples|: Integer specifying the number of bootstrap samples to generate. The default is 0.
	\item \verb|parallel|: Logical scalar indicating whether to parallelize simulations of different interventions to multiple cores, as well as whether to parallelize bootstrapping.
	\item \verb|ncores|: Integer specifying the number of CPU cores to use in parallel simulation. This argument is required when parallel is set to \verb|TRUE|. In many applications, users may wish to set this argument equal to \verb|parallel::detectCores() - 1|.
\end{itemize}

\section{Discussion}\label{discussion}

The \pkg{gfoRmula} package provides an implementation of the parametric g-formula in \proglang{R} for estimating the effects of time-varying treatment strategies from longitudinal data with time-varying confounding. The package handles survival outcomes with or without competing events, as well as end of follow-up outcomes.  It provides flexible options for estimating the required conditional covariate distributions and outcome means in realistic settings with many follow-up times and high-dimensional confounders.  It allows for joint interventions on multiple time-varying treatments, flexible intervention rules and the ability to incorporate \textsl{a priori} deterministic knowledge of covariate distributions when available. 

The parametric g-formula is only one of several estimators that may be used to estimate any of the various g-formula functions considered in Section \ref{assumptions} in high-dimensional settings. Other approaches include iterative expectation estimators \citep{Bangandrobins}, simple inverse probability weighted estimators \citep{coxmsm,msmref,msmjr} or doubly-robust approaches \citep{Bangandrobins,ltmle}.  These different approaches (while nonparametrically equivalent in the sense that they will give equivalent results when all nuisance parameters can be estimated in the absence of any model constraints) may give different results in realistic settings where some constraints on the nuisance parameters (here, the conditional covariate distributions and outcome means) are required. 

These different methods will have different advantages and disadvantages in practice.  A key advantage of the parametric g-formula is the computational ease under which many different causal questions may be considered in one function call.   This is at the price, however of a relatively greater potential for model misspecification bias, particularly when there are many timepoints and many time-varying confounders.  In fact, it has been shown that at least some model misspecification bias can be guaranteed in parametric g-formula estimates when (i) the null is true, (ii) standard parametric models are used for estimating the conditional covariate distributions and outcome means and (iii) time-varying confounders are affected by past treatment \citep{gnull}.  

However, the relative performance of the parametric g-formula compared to the other methods cited above has not been thoroughly studied, particularly given the ability to incorporate knowledge of covariate distributions when available (which cannot be incorporated into the alternative estimators noted above) or when nonparametric estimation algorithms are used. In addition to providing a flexible tool for estimating effects of time-varying treatment strategies in observational data, the \pkg{gfoRmula} package can facilitate simulation studies of the performance of this method in different settings, particularly relative to other methods currently only implemented in \proglang{R}.

\section{Acknowledgements}
The authors thank L.Paloma Rojas Saunero, Eleanor Murray and Sara Lodi for testing of the package.  This work was funded by National Institutes of Health grant NIH R37 AI102634.

\bibliography{refs}

\end{document}